\documentclass[10pt,superscriptaddress,twocolumn,aps,showpacs,nofootinbib,eqsecnum]{revtex4}

\usepackage[pdfborder={0 0 0},colorlinks=true,linkcolor=blue,citecolor=blue %
,filecolor=blue,urlcolor=blue]{hyperref}   
\usepackage{url}
\usepackage{ulem}
\usepackage{slashed}
\usepackage{amssymb}
\usepackage{amsfonts}
\usepackage{mathbbol} 
\usepackage{amstext}
\usepackage{graphicx}
\usepackage{color}
\usepackage{epsfig,amsmath,lscape}

\newcommand{\myindent}{\hspace{15pt}}

\begin{document}

\title{Hadron-quark phase transition in asymmetric matter with boson condensation}

\author{Rafael Cavagnoli} %\email{rafael@fsc.ufsc.br }
\affiliation{Departamento de F\'{\i}sica - CFM - Universidade Federal de Santa Catarina, \\ 
Florian\'opolis - SC - CP. 476 - CEP 88.040 - 900 - Brazil}
\affiliation{Centro de F\'{i}sica Te\'{o}rica, Departamento de F\'{i}sica,
Universidade de Coimbra, P-3004-516 Coimbra, Portugal}

\author{Constan\c{c}a Provid\^encia}
\affiliation{Centro de F\'{i}sica Te\'{o}rica, Departamento de F\'{i}sica,
Universidade de Coimbra, P-3004-516 Coimbra, Portugal}

\author{Debora P. Menezes}
\affiliation{Departamento de F\'{\i}sica - CFM - Universidade Federal de Santa Catarina, \\ 
Florian\'opolis - SC - CP. 476 - CEP 88.040 - 900 - Brazil}

\begin{abstract}

In the present work we study the hadron-quark phase transition with boson 
condensation in asymmetric matter by investigating the binodal surface and 
extending it to finite 
temperature in order to mimic the QCD phase diagram. 
We consider a system with two conserved charges (isospin and baryon densities) 
using the Gibbs' criteria for phase equilibrium. 
In order to obtain these conditions we use two different models for the two 
possible phases, namely the non-linear Walecka model (NLWM) for the 
hadron matter (also including hyperons) and the MIT bag model for 
the quark phase. It is shown that the phase transition is very  sensitive to the  density dependence of the equation of
state and the symmetry energy. For isospin asymmetry of 0.2 and a mixed phase with a fraction
of 20\% of quarks, a transition density in the interval $2\rho_0<\rho_t< 4\rho_0$ was obtained
for temperatures $30<T<65$ MeV. 

\end{abstract}


\maketitle

%%%%%%%%%%%%%%%%%%%%%%%%%%%%%%%%%%%%%%%%%%%%%
\section{Introduction}

\myindent Since some decades ago just after the discovery of asymptotic freedom of QCD 
\cite{asymp-freedom} the possibility of the existence of a new state of matter in 
high energy physics is under consideration, namely, a color deconfined phase 
of quarks and gluons, the so-called quark-gluon plasma (QGP) \cite{qgp-refs}. The 
main goal of the heavy-ion collision experiments at ultra-relativistic energies 
is to create, under controlled conditions, and understand the properties of this 
new state of matter. This opens a new field of study in strong interaction physics.

\myindent Many experiments have been proposed and accelerators built in the search for 
the QGP at different energies at SIS/GSI, AGS/BNL, SPS/CERN, RHIC/BNL and 
LHC/CERN to look for some signs and signatures of the production of the QGP that 
subsequently hadronizes \cite{exp}. The study of particle production in ion collisions 
contribute to the understanding of the conditions under which the quark-gluon plasma 
may be produced and also to determine the equations of state (EoS) of strongly 
interacting matter.

\myindent In hydrodynamical models the system which arises from a high-energy 
collision (fireball) reaches an approximately local thermal equilibrium 
(thermalization) {and } expands evolving collectively up to the point when the mean 
free path of the created and interacting particles becomes 
large enough for the particles to escape from the fluid, 
i.e., the interactions among the particles of the system cease because 
the system has reached the freeze out point. 
The approximately local thermalization is considered due to detailed 
computations of the expansion stage that takes much longer than 
the typical scattering times \cite{thermalization-mclenan}. Although, 
non-equilibrium processes are also important for the dynamics, equilibrium 
processes are a quite good approximation to be used in theoretical models, and are 
a reasonable first approximation at 
freeze-out. Some authors consider { both } the temperature at which 
the inelastic collisions cease (chemical freeze-out) and the elastic 
collisions cease (kinetic freeze-out) \cite{ch-kt-freezeout}.

\myindent In the end of the eighties of the last century the existence of the 
(chiral) critical end point (CEP) in the QCD  phase diagram was suggested 
\cite{cep-asakawa,cep-barducci} and since then its properties have been extensively 
studied \cite{cep-studied}.Although, most lattice QCD calculations indicate the 
existence of the CEP for $\mu_B > 160$~MeV \cite{cep-lattice1,cep-lattice2,cep-lattice3}, 
its exact location is not well known since it depends, for example, on the mass of 
the strange quark. The CEP separates second-order transition at high temperatures (or 
even a smooth cross-over) from the first-order transition at high chemical potentials 
in the QCD phase diagram. Studying this intermediate region is a hard task since 
perturbation theory cannot be applied to QCD at this regime and additionally at 
finite chemical potential the usual lattice approach fails. Moreover, new techniques 
have been proposed to study lattice QCD at finite T and $\mu$ \cite{lattice-qcd-t}.
On the other hand the lattice QCD simulations of different groups disagree with each 
other on the location of the critical end point.

\myindent Subsequently in the late nineties a hypothesis arose \cite{onset-gorenstein} 
that the onset of the deconfinement phase transition was located between the top AGS 
and SPS energies. The CERN energy scan program of the NA49 experiment at SPS has given 
signs of a phase change at $E_{\rm lab} \sim 30$~A$\cdot$GeV particularly from the 
horn-like peak in the $K^+ / \pi^+$ ratio \cite{sps-scan-na49}. 
%
%%%%%%%%%%%%%%%%%%%%%%%%%%%%
%\begin{widetext}
%
%\begin{center}
 \begin{table*}[ht]
\centering
\begin{tabular}{cccccccc}
\hline \hline
 & ~ SIS/GSI ~ & ~ Synchr./JINR ~ & ~ AGS/BNL ~ & $\downarrow$ & ~ SPS/CERN ~ & ~ RHIC/BNL~ & ~ LHC/CERN ~ \\ 
\hline
 &  &  &  &  &  &  &  \\ 
~ $E_{\rm lab}$~(A$\cdot$GeV)  ~ & 2.0 & 4.2 & 14.6 &  & 158 & 2.1~$\times 10^4$ & 1.6~$\times 10^7$ \\ 
~ $\sqrt{s_{\scriptscriptstyle NN}}$~(GeV) ~ & 2.7 & 3.4 & 5.6 &  & 17.3 & 200 & 5400 \\ 
%\hline
% &  &  &  & $\swarrow \searrow$ &  &  &  \\ 
\hline \hline
 &  &  & FAIR/GSI
 &  & NICA/JINR
 &  &  \\ 
\hline
 &  &  &  &  &  &  &  \\ 
~ $E_{\rm lab}$~(A$\cdot$GeV)  ~ &  &  & 34 &  & 40 &  \multicolumn{2}{c}{($\leftarrow$ planned facilities)}  \\ 
~ $\sqrt{s_{\scriptscriptstyle NN}}$~(GeV) ~ &  &  & 8.2 &  & 9 &  &  \\ 
\hline \hline
\end{tabular}
\label{tab-energies}
\caption{Ion beam top energies in some collision experiments.} 
\end{table*} 
%
%\end{center}
%
%\end{widetext}
%%%%%%%%%%%%%%%%%%%%%%%%%%%%

\myindent Furthermore, the hadronic freeze-out estimated for different colliding energies 
\cite{mfreeze-cley} shows a maximum at $\sqrt{s_{\scriptscriptstyle NN}}= 4+4$~GeV, which can be 
reached for a fixed-target bombarding energy of $20-30$~A$\cdot$GeV at the baryonic chemical 
potential region $\mu_B = 400-500$~MeV. In addition, hydrodynamical calculations 
\cite{hydro-ivanov,hydro-skokov} of phase trajectories during collisions, in the 
QCD phase diagram, indicate that for 
$E_{\rm lab} \sim 30$~A$\cdot$GeV ($\sqrt{s_{\scriptscriptstyle NN}} \sim 8$~GeV) the trajectory goes near 
the CEP. 

\myindent Since then, the interest on the intermediate energies (not ultra-relativistic) in 
collision experiments is increasing as well as the theoretical study of the phase 
transition at that regime. 

\myindent For this purpose the new facilities, namely, the \textbf{N}uclotron-based 
\textbf{I}on \textbf{C}ollider 
f\textbf{A}cility (NICA) at JINR/ Dubna \cite{nica-jinr} and the \textbf{F}acility for 
\textbf{A}ntiproton and \textbf{I}on \textbf{R}esearch (FAIR) at GSI/Darmstadt 
\cite{fair-gsi}, give the opportunity to explore 
an interesting region of the phase diagram, in search %of 
 for the QGP and where the critical end point (CEP) is expected 
to exist, complementing other heavy-ion collision 
experiments (NA61-SHINE, low-energy RHIC) compatible with the energy range 
$E_{\rm lab} = 2-40$~A$\cdot$GeV. It is also important to consider the possibility that 
new features arise at still lower temperatures and higher densities as the color 
superconducting quark matter phases like the 2SC phase and the recently conjectured 
quarkyonic phase \cite{quarkyonic}. A summary of ion beam top energies used in some 
collision experiments is shown in table \ref{tab-energies}.

\myindent It is also possible to study the phase transition from hadronic matter 
to a quark phase within the effective models that describe two separated phases, 
and also the structure of the mixed phase can be obtained through the Gibbs' 
conditions \cite{landau-gibbs}. Some features of this phase transition can be 
obtained by means of the binodal surface, which is a phase coexistence curve 
in the parameter space. 

\myindent The different ion beams used in collision experiments present different numbers 
of neutrons (N) and protons (Z). It is also interesting to study the isospin effects 
on the transition to a mixed phase of hadrons and quarks. We can define the asymmetry 
parameter (isospin ratio) of a nucleus (or the hadron phase) as $\alpha \equiv (N - Z)/(N + Z)$, 
such that, $\alpha$ runs from 0 (symmetric matter) to 1 (pure neutron matter). From 
table \ref{ion-asymmetries}\ref{ion-asymmetries} one sees some ions used in nucleus-nucleus 
collisions and the respective asymmetry parameter of each system. Systems with 
isospin ratios $0 \leq \alpha \leq 0.23$ are up to now experimentally accessible in 
ion collisions and the case $\alpha = 1.0$ corresponds to neutron matter 
which is relevant in some astrophysical applications.

\myindent We study the phase transition from hadrons to a quark-gluon plasma in 
asymmetric matter using a two-phase model, analyzing the features that 
depend on the isospin and may be relevant in a phenomenological description of 
heavy-ion collisions \cite{muller-serot,muller-pions,ditoro06,ditoro2010}.  

\myindent  It is interesting to investigate asymmetric systems since in the 
liquid-gas phase transition of nuclear matter the asymmetric case shows 
different properties from the symmetric one 
\cite{muller-serot,barranco-sym,glen-sym}. It is shown that 
the transition of an asymmetric system is of second order (continuous) rather 
than the first order (discontinuous) transition in symmetric systems 
\cite{muller-serot,muller-pions,ditoro06,ditoro2010}. 

\myindent Hence, an interesting task is to investigate the isospin effect on 
the hadron-quark phase transition at lower temperatures and densities higher 
than the saturation density of the normal nuclear matter, that can be probed 
in heavy-ion collisions at intermediate energies. In addiction, the presence 
of bosons can  modify the isospin of the hadron phase. Also, at low 
temperatures these features depend strongly on the nuclear symmetry energy. On 
the other hand, 
at higher temperatures the inclusion of bosons shows an interesting feature due 
to the onset of a boson condensate in asymmetric systems if we consider an 
approximately local thermal equilibrium.

\begin{table}[ht]
\begin{tabular}{|c|c|c|c|}
\hline
 & ~$^{12}$C~$+$~$^{12}$C~ & ~$^{20}$Ne~$+$~$^{20}$Ne~ & ~$^{58}$Ni~$+$~$^{58}$Ni~     \\ 
\hline
~~ $\alpha$ ~~ & 0 & 0 & 0.034    \\ 
\hline
\multicolumn{1}{l}{} & \multicolumn{1}{l}{} & \multicolumn{1}{l}{} & \multicolumn{1}{l}{}  \\
\hline
 & ~$^{20}$Ne~$+$~$^{63}$Cu~ & ~$^{20}$Ne~$+$~$^{118}$Sn~ & ~$^{118}$Sn~$+$~$^{118}$Sn~   \\ 
\hline
~~ $\alpha$ ~~ & 0.060 & 0.130 & 0.150   \\ 
\hline
\multicolumn{1}{l}{} & \multicolumn{1}{l}{} & \multicolumn{1}{l}{} & \multicolumn{1}{l}{}  \\
\hline
 & ~$^{20}$Ne~$+$~$^{209}$Bi~ & ~$^{197}$Au~$+$~$^{197}$Au~ & ~$^{20}$Ne~$+$~$^{238}$U~    \\ 
\hline
~ $\alpha$ ~ & 0.188 & 0.198 & 0.201   \\ 
\hline
\multicolumn{1}{l}{} & \multicolumn{1}{l}{} & \multicolumn{1}{l}{} & \multicolumn{1}{l}{}  \\
\hline
 & ~$^{197}$Au~$+$~$^{208}$Pb~ & ~$^{208}$Pb~$+$~$^{208}$Pb~ & ~$^{238}$U~$+$~$^{238}$U~   \\ 
\hline
~ $\alpha$ ~ & 0.205 & 0.211 & 0.227   \\ 
\hline
\end{tabular}
\label{ion-asymmetries}
\caption{Some ions used in collision experiments and the respective asymmetry 
parameter ($\alpha$) of the system.}
\end{table} 

\myindent This approach is useful for providing a qualitative orientation on the 
features that arise when a phase transition from hadrons to quarks takes place and 
two conserved charges are considered, i.e. at finite baryon density and isospin. 

\myindent As already mentioned, the problem we investigate in the present 
paper has already been studied in previous works \cite{muller-pions,ditoro06,
ditoro2010} within different perspectives, based on different parametrizations 
and containing different ingredients. In \cite{muller-pions} the hadronic phase 
is given by one parametrization of the non-linear Walecka model, the quark
phase is calculated with the MIT bag model for one specific value of the bag 
constant and pions are included. In \cite{ditoro06,ditoro2010} a mixed phase 
of hadrons and quarks is particularly emphasized  and the influence of the 
symmetry energy on 
the phase transition investigated.  In  \cite{ditoro2010} neither bosons nor gluons 
were considered and the quark phase was described within the MIT bag model, and in 
\cite{ditoro06} two quark different models have been used: the MIT bag
model (with and without gluons) and the color dielectric quark model. In these 
works five different parametrizations of the 
non-linear Walecka model are considered and one of them includes the $\delta$ 
mesons. All these models have a quite high value of the symmetry energy slope, namely
$85<L<103$ MeV, and therefore they have a quite hard symmetry energy at 
intermediate densites, the densities of interest for the present work.

\myindent In the present work we consider seven different parametrizations of
the non-linear Walecka model, which span  a large variety of EOS: they include hard and soft
EOS with hard and soft symmetry energies at intermediate densities. This will allow us the see
the effect of both the isoscalar   and the isovector interaction on the phase
transition. For the quark model we have considered the MIT bag model with  various values of
the bag constant and with gluons.
The bag constant was chosen in accordance with heavy-ion collision data.
 In the hadronic phase we have  studied the effect of including two kinds 
of bosons, pions and kaons. This more complete 
picture allows us to discuss various aspects of the phase transition at finite 
temperature which were not discussed before.

\myindent The remainder of this article is organized as follows: In section II we 
present the formalism used in this work. In section III the mixed phase features are 
presented and in section IV we show the numerical results and discussion. Finally, in 
section V we summarize the results and give a brief concluding discussion.

%%%%%%%%%%%%%%%%%%%%%%%%%%%%%%%%%%%%%%%%%%%%%
\section{The Formalism}

\myindent In the present section we present the equations of state (EoS) for the hadron phase 
and for the quark phase used in this work and their respective definitions. Bosons 
are included using a meson-exchange 
type Lagrangian that couples the bosons to meson fields and the possibility of a boson 
condensate is also presented.

%
%
%%%%%%%%%%%%%%%%%%%%%%%%%%%%%%
\begin{center}
%\medskip
\textbf{A. Quark phase: quarks u, d (+ gluons)}
\medskip
\end{center}

\myindent Quark matter has been extensively described by the MIT bag model \cite{bag}.
In its simplest form, the quarks are considered to be free inside a Bag and 
the thermodynamic properties are derived from the Fermi gas model in two 
limits: $T = 0$, $m_q \neq 0$ and $T \neq 0$, $m_q = 0$. The energy density, 
the pressure and the quark $q$ density are respectively given by:
\begin{equation}
{\cal E}_q= 3 \times 2  \sum_{q=u,d} \int \frac{d^3p}{(2\pi)^3}
\sqrt{{\mathbf p}^2+ m_q^2} \left(f_{q+}+f_{q-}\right) + B \,,
\label{ener-quark}
\end{equation}
\begin{equation}
P_q =\frac{1}{\pi^2} \sum_{q}
\int d p \frac{{\mathbf p}^4}{\sqrt{{\mathbf p}^2+m_q^2}} 
\left(f_{q+} + f_{q-}\right) - B \,,
\label{press-quark}
\end{equation}
\begin{equation}
n_q = 3 \times 2 \int\frac{d^3p}{(2\pi)^3}(f_{q+}-f_{q-}) \,,
\label{rhoq}
\end{equation}
\begin{equation}
f_{q\pm}=\frac{1}{1 + e^{[(\epsilon_q \mp \mu_q)/T]} }  \,, 
\end{equation}

where $3$ stands for the number of colors, $2$ for the spin degeneracy,
$m_q$ for the quark masses, $B$ represents the bag
pressure and $f_{q\pm}$ the distribution functions for the quarks and anti-quarks, 
$\epsilon_q = \sqrt{{\mathbf p}_q^2+m_q^2}$, $\pm \mu_q$ being the chemical potential for 
quarks and anti-quarks of type $q$,
 \begin{equation}
\mu_u=\frac{2 \mu_p-\mu_n}{3} \, , \quad \mu_d=\frac{2 \mu_n-\mu_p}{3} \;.
\end{equation}
$ $

The quark density is

\begin{equation}
\quad n_{q} = n_u + n_d \;,
\end{equation}
$ $

and the ``quark baryon density'' is given by:

 \begin{equation}
\quad n_{B}^Q = \frac{n_u + n_d}{3} \;.
\end{equation}
$ $

The thermodynamic potential per unit volume of the MIT bag model 
(two-flavor case) and the corresponding equations of state (EoS) \cite{muller-pions,gle1}
for massless quarks and a Bose gas of gluons of degeneracy $\gamma_g = 2 \times 8$ 
with the lowest-order gluon interaction ($\alpha_s$) is 
\begin{widetext}

\begin{equation}
\frac{\Omega_{\scriptscriptstyle QGP}}{V} = - \frac{\pi^4}{45} T^4 \left( 8 + \frac{21}{4}N_f \right) 
- \frac{1}{2} \sum_{q=u,d}^{} \left( T^2 \mu_q^2 + \frac{\mu_q^4}{2 \pi^2} \right)  
+ \frac{2 \pi}{9} \alpha_s \left[ T^4 \left( 3 + \frac{5}{4}N_f \right) 
+ \frac{9}{2} \sum_{q=u,d}^{} \left( \frac{T^2 \mu_q^2}{\pi^2} + \frac{\mu_q^4}{2 \pi^4} \right) \right] + B \;, 
\label{pot-mac-qgp}
\end{equation}
$ $

from where we can obtain the pressure $P_{\scriptscriptstyle QGP} = - \Omega_{\scriptscriptstyle QGP}/V$ 
 the energy density and the quark number density:

\begin{equation}
P_{\scriptscriptstyle QGP} = \frac{8 \pi^2}{45}T^4 \left( 1 - \frac{15 \alpha_s}{4 \pi} \right) + 
\sum_{q} \left[ \frac{7}{60}\pi^2 T^4 \left( 1 - \frac{50 \alpha_s}{21 \pi} \right) 
+ \left( \frac{1}{2}T^2 \mu_q^2 + \frac{1}{4 \pi^2 }\mu_q^4 \right) \left( 1 - \frac{2 \alpha_s}{\pi} \right)   \right] - B  \;,
\end{equation}

\begin{equation}
{\cal E}_{\scriptscriptstyle QGP} = 3P_{\scriptscriptstyle QGP} + 4B \qquad \qquad {\rm ;} \qquad \qquad
n_q = \sum_{q}^{} \left( T^2 \mu_q+ \frac{\mu_q^3}{\pi^2} \right) \left( 1 - \frac{2 \alpha_s}{\pi} \right)  \;.
\label{energ-dens-qgp}
\end{equation}

\end{widetext}

The strong coupling $\alpha_s$ is taken as a constant in the 
present work ($\alpha_s = 0.349$) and $N_f$ stands for the number of flavors 
($N_f = 2$, quarks $u$ and $d$).

%
%%%%%%%%%%%%%%%%%%%%%%%%%%%%%%%%%%
\begin{center}
\textbf{B. Hadron phase:  nucleons (+ hyperons)}
\medskip
\end{center}

\myindent The equations of state of asymmetric matter
within the framework of the relativistic non-linear Walecka model (NLWM)
\cite{bb} are presented next. In this  model the nucleons are coupled to 
neutral scalar $\sigma$, 
isoscalar-vector $\omega^\mu$ and isovector-vector $\vec \rho^\mu$  meson fields. 
The Lagrangian density reads

%\begin{widetext}
%\begin{equation}
\begin{eqnarray} 
&{\cal L}_{B}&= {\bar \psi \left[ ~ { \gamma_\mu  \left( {i \partial^\mu 
- g_{\omega j} \, \omega^\mu - g_{\rho j} \, \vec \tau _j \,.\, \vec \rho^{\, \mu}  } \right) 
- m_j^* ~} \right]\psi }  \nonumber \\ 
&+& \frac{1}{2} {\partial_\mu \sigma \partial^\mu \sigma - \frac{1}{2} m_{\sigma}^2 \sigma^2 }  
- \frac{1}{{3!}} k \sigma^3  - \frac{1}{{4!}} \lambda \sigma^4     \nonumber \\
&-& \frac{1}{4} \Omega_{\mu \nu} \, \Omega^{\mu \nu}
+ \frac{1}{2} m_{\omega}^2 \, \omega_\mu \omega^\mu 
+ \frac{1}{4!}\xi g_{\omega}^4 (\omega_{\mu} \omega^{\mu})^2  \nonumber \\
&-& \frac{1}{4} \vec R_{\mu \nu } \,.\,  \vec R^{\mu \nu } + 
\frac{1}{2} m_\rho^2 \, \vec \rho_{\mu} \,.\, \vec \rho^{\, \mu}   \nonumber \\
&+& \Lambda_{\rm v} (g_{\rho}^2 \; \vec \rho_{\mu} \,.\, \vec \rho^{\, \mu} )(g_{\omega}^2 \; \omega_{\mu} \omega^{\mu} )  \;,   %\\
%\nonumber \\
%
%&  \;, & \nonumber
\label{baryon-lag}   
\end{eqnarray}
%
%\end{equation}
%\end{widetext}
$ $

where $m_j^* = m_j - g_{\sigma j}\, \sigma$ is the baryon effective mass,  
$\Omega_{\mu\nu }=\partial_\mu \omega_\nu - \partial_\nu \omega_\mu$~, 
$\vec R_{\mu \nu } = \partial_\mu \vec \rho_\nu -\partial_\nu \vec \rho_\mu 
-g_\rho \left({\vec \rho_\mu \,\times \, \vec \rho_\nu } \right)$, $g_{ij}$ 
are the coupling constants of mesons $i = \sigma, \omega, \rho$ with baryon 
$j$, and $m_i$ is the mass of meson $i$. The couplings 
$k$~($k = 2\,M_N\,g_{\sigma}^3\,b$) and $\lambda$~($\lambda = 6\, g_{\sigma}^4\,c$) 
are the weights of the non-linear scalar terms and $\vec \tau$ is the isospin operator. 
This Lagrangian includes an isoscalar-isovector mixing term 
$\Lambda_{\rm v} (g_{\rho}^2 \; \vec \rho_{\mu} \,.\, \vec \rho^{\, \mu} )(g_{\omega}^2 \; \omega_{\mu} \omega^{\mu} )$ 
as presented in \cite{piekarewicz1} which plays an important role 
in high densities.
It can also be extended to include all the hyperons from the baryon octet. 

\myindent Within the relativistic mean field (RMF) framework the thermodynamic 
potential per unit volume corresponding to the Lagrangian density 
(\ref{baryon-lag}) is

\begin{eqnarray} 
\displaystyle \frac{\Omega_B}{V} &=& \frac{1}{2} m_{\sigma}^2 \sigma_0^2 
+ \frac{1}{3!} k \sigma_0^3 + \frac{1}{4!} \lambda \sigma_0^4 
- \frac{1}{2} m_{\omega}^2 \omega_0^2 - \frac{1}{4!} \xi \omega_0^4    \nonumber \\
\nonumber \\
&-& \frac{1}{2} m_{\rho}^2 \rho_{03}^2 -  2T \sum_{j} \int \frac{d^3 p}{(2\pi)^3} \left\{  
\ln \left[ 1 + e^{-\beta(E_j^* - \nu_j)} \right]  \right.  \nonumber \\
\nonumber \\
&+& \left. \ln \left[ 1 + e^{-\beta(E_j^* + \nu_j)}  \right]  \right\} 
- \Lambda_{\rm v} \, g_{\rho}^2 \, g_{\omega}^2 \, \omega_0^2 \, \rho_{03}^2    \;,  
\label{therm-pot-baryons}
\end{eqnarray}
$ $

where $\beta = 1/T$, $E_j^* = ({\mathbf p}_j^2 + M_j^{*\,2})^{1/2}$ 
and the effective chemical potential of baryon $j$ is given by
\begin{equation}
  \nu_j = \mu_j - g_{\omega}\omega_0 - \tau_{3j}  \, g_{\rho}\, \rho_{03} \;.
\label{pot-quim-nucleons}
\end{equation}

The EoS for the baryons can then be calculated

\begin{align} 
P_{B} &= \frac{1}{3 \pi^2} {\sum\limits_{j}^{} {\int_{}^{ }} 
{\frac{p^4 dp}{\sqrt {p^2 + m_j^{*2}}}}} (f_{Fj+} \,+\, f_{Fj-}) 
+ \frac{{m_\omega^2 }}{2} \omega_0^2  \qquad \nonumber \\
&+ \frac{\xi}{24} \omega_0^4 + \frac{{m_\rho ^2 }}{2} \rho_{03}^2 
- \frac{m_\sigma^2}{2} \sigma_{0}^2 - \frac{k}{6} \sigma _{0}^3 
- \frac{\lambda }{24} \sigma _{0}^4  \qquad \nonumber \\
\nonumber \\
&+ \Lambda_{\rm v} \, g_{\rho}^2 \, g_{\omega}^2 \, \omega_0^2 \, \rho_{03}^2  \;,
\label{press-bar}
\end{align} 

\begin{align} 
{\cal E}_{B} &= \frac{1}{\pi^2}  {\sum\limits_{j}^{} {\int_{}^{} {p^2 dp \sqrt {p^2 + m_j^{*2}}}} (f_{Fj+} \,+\, f_{Fj-}) }  
+ \frac{{m_\omega^2 }}{2} \omega_0^2  \quad \nonumber \\
&+ \frac{\xi}{8} \omega_0^4 + \frac{{m_\rho ^2 }}{2} \rho_{03}^2 
+ \frac{{m_\sigma^2 }}{2} \sigma _{0}^2 + \frac{k}{6} \sigma _{0}^3 
+ \frac{\lambda }{{24}} \sigma _{0}^4  \quad \nonumber \\
%\nonumber \\
%
&+ 3 \Lambda_{\rm v} \, g_{\rho}^2 \, g_{\omega}^2 \, \omega_0^2 \, \rho_{03}^2  \;,
\label{deng-bar}
\end{align}
\begin{equation}
n_{B}^{{\kern 2pt}j} = \frac{2}{(2 \pi)^3} \int\limits_{}^{} {(f_{Fj+} - f_{Fj-})\,d^3 p }  \;.
\label{rhoi2-octeto}
\end{equation}

and $f_{Fj\pm}$ is the Fermi distribution for the baryon (+) and anti-baryon (-) $j$:
\begin{equation}
f_{Fj\pm} = \frac{1}{e^{\beta(E_j^* \mp \nu_j)} + 1} .
\end{equation}

%%%%%%%%%%%%%%%%%%%%%%%%%%%%%%%
\begin{center}
\medskip
\textbf{C. Hadron phase: bosons (pions + kaons)}
\medskip
\end{center}

\myindent It is possible to include the boson fields using terms from the 
chiral 
perturbation theory \cite{nelson-kaplan1}. In the present work we prefer to use 
a meson-exchange 
type Lagrangian that couples the bosons to meson fields. For simplicity we 
apply the same approach to the kaons and pions. The Lagrangian density 
in the minimal coupling scheme \cite{knorren-lag,schaffner-lag,prakash-rep-lag,glen-shaffner-lag,pons1,kapusta-lag} 
is given by:
\begin{equation}
{\cal L}_{b} = D_{\mu}^* \, \Phi^* \,\, D^{\mu} \, {\Phi} - m_{b}^{*2} \Phi^* \, {\Phi} \;,
\label{bosons-lag}
\end{equation}
where the covariant derivative is:

\begin{equation}
D_{\mu} = \partial_{\mu} + i X_{\mu} \;,
\label{covar-deriv}
\end{equation}
\begin{equation}
X_{\mu} \equiv g_{\omega b}~ \omega_{\mu} + g_{\rho b}~ \vec{\tau}_b \cdot \vec \rho_{\mu} \;,
\end{equation}
$ $
and the boson effective mass, $m_{b}^{*} = m_{b} - g_{\sigma b}~ \sigma$.~ 
The boson field can then represent either the kaons or pions (particles and 
anti-particles):
\begin{equation}
\Phi \equiv (K^+, K^0) \quad , \quad  \Phi^* \equiv (K^-, \bar{K}^0) \;,
\end{equation}
or
\begin{equation}
\Phi \equiv (\pi^-,\pi^0) \quad , \quad  \Phi^* \equiv (\pi^+,\pi^0) \;.
\end{equation}

The isospin third-component to the bosons is given by
\begin{equation}
 \tau_{3 \pi} = \left\{ {\begin{array}{l}
   { + 1 \;\; , \;\; \pi^+ }  \\
  \;\; 0 \;\; \; , \;\; \pi^0 \\
   { - 1 \;\; , \;\; \pi^- }  \\
\end{array}} \right. ; \quad 
\tau_{3 K} = \left\{ {\begin{array}{l}
   { + \frac{1}{2} \; , \; K^+ , \bar{K}^0}  \\
\\
   { - \frac{1}{2} \; , \; K^- , K^0}  \\
\end{array}} \right.
\end{equation}
In order to obtain the boson thermodynamic potential it is also possible to 
perform a similar calculation as carried out in reference \cite{kapusta-lag} with 
the respective modifications in the covariant derivative (\ref{covar-deriv}). 

\myindent As the neutral pion is its own anti-particle we need to set $X_{\mu} = 0$~
($g_{\omega \pi^0}=g_{\rho \pi^0}=0$, and if required $m_{\pi^0}^* = m_{\pi^0}$, 
then $g_{\sigma \pi^0}=0$) to achieve the 
correct thermodynamical features of these uncharged particles. In this case, the Lagrangian 
(\ref{bosons-lag}) takes up its simplest form and $\Phi^* = \Phi$. In particular, 
$\pi^0$ results completely decoupled from the other particles and its population 
is that of a boson gas at temperature $T$ and $\mu_{\pi^0} = 0$.

\myindent In the Appendix we show the calculation of the bosonic EoS:

\begin{widetext}

\begin{equation}
P_b = \zeta^2 \left[ (\mu_b - X_0 )^2 - m_b^{*\,2} \right] 
- T \int \frac{d^3 p}{(2 \pi)^3}  \left\{ \; \ln \left[ 1 - e^{-\beta(\omega^+ - \mu)}  \right] 
+  \ln \left[ 1 - e^{-\beta(\omega^- + \mu)}   \right] \; \right\} \;,
\label{boson-press}
\end{equation}

\begin{equation}
{\cal E}_b = \zeta^2 \left[ m_b^{*\,2} + \mu_b^2 - X_0^2 ) \right] 
+ \int \frac{d^3 p}{(2 \pi)^3}  \left\{ \;  \omega^+ f_{B+}   
\; + \;   \omega^- f_{B-}    \; 
\mathop {}\limits_{} \right\} \;,
\label{boson-energ}
\end{equation}

\begin{equation}
n_b = 2 \zeta^2 (\mu_b - X_0) + \int \frac{d^3 p}{(2 \pi)^3}  \left\{ \; f_{B+}  
\; -  \; f_{B-}    \; 
\mathop {}\limits_{}  \right\} \;,
\label{boson-dens}
\end{equation}
$ $

where the Bose distribution for particles ($f_{B+}$) and anti-particles ($f_{B-}$)
 appears naturally in the EoS and read: 

\begin{equation}
f_{B \pm} = \frac{1}{e^{\beta(\omega^{\pm} \mp \mu_b)} - 1}  
= \frac{1}{e^{\beta \left[  (\epsilon_b^* \pm X_0) \mp \mu_b \right]} - 1}  
= \frac{1}{e^{\beta  (\epsilon_b^*  \mp \nu_b )} - 1} \;,
\end{equation}

with $\epsilon_b^* = \sqrt{p^2 + m_b^{*\,2}}$~, and hence we define the boson 
effective chemical potential as
\begin{equation}
\nu_b \equiv \mu_b - X_0 \;.
\label{nu}
\end{equation}

From equation (\ref{boson-dens}) one notes two contributions in the boson density and 
we can define them as the ``condensate'' and ``thermal'' ones
\begin{equation}
n_b = n_b^c(\zeta) + n_b^T(T)  \;,
\end{equation}
$ $

and the entropy density is given by $s_b = \beta (P_b + {\cal E}_b - \mu_b n_b)$~. 
The order parameter $\zeta$ can be obtained through  the minimization of the thermodynamic 
potential.

\end{widetext}

%%%%%%%%%%%%%%%%%%%%%%%%%%%%%%%
\begin{center}
\medskip
\textbf{D. Hadron phase equations}
\medskip
\end{center}

The thermodynamic potential of the hadron phase (HP) including both the baryons and 
the bosons, is given by 
\begin{equation}
\Omega_{HP} = \Omega_{B} + \Omega_{b}
\label{therm-pot-hp}
\end{equation}

where $\Omega_{B}$ is given by (\ref{therm-pot-baryons}) and $\Omega_{b}$ by 
(\ref{therm-pot-boson}). By minimizing the thermodynamic potential $\Omega_{HP}$ 
with respect to the meson fields $\sigma$, $\omega$ and $\rho$, and also with respect  
to the order parameter $\zeta$, within the mean-field approximation 
($\sigma \to \langle \sigma \rangle = \sigma_{0}~ ;\;
\omega_\mu \to \langle \omega_\mu  \rangle = \delta_{\mu 0} \,\omega_0 ~;\; 
\vec \rho_\mu  \to \langle \vec \rho_\mu  \rangle = \delta_{\mu 0} \, \delta^{i 3} \rho_0^3  
\equiv \delta_{\mu 0} \, \delta^{i 3} \rho_{03}  $), we obtain the equations for the 
hadron phase:

\begin{equation}
m_{\sigma}^2 \, \sigma_{0} = - \frac{k}{2} \sigma _{0}^2 - \frac{\lambda }{6} \sigma _{0}^3 
+ \sum\limits_j^{} g_{\sigma j} \, n_{j}^{s} + \sum\limits_{b}^{} g_{\sigma b} ( n_b^c  + n_b^s)  \;, \nonumber
\label{sigma-field-boson1}
\end{equation}
\begin{align} 
m_{\omega}^2 \, \omega_{0} &= -\dfrac{\xi g_{\omega}^4}{6} \omega_0^3 
+ \sum\limits_j^{} g_{\omega j} \, n_{j}^{} 
+ \sum\limits_{b}^{} g_{\omega b} \, n_b
  \quad \nonumber \\
&- 2 \Lambda_{\rm v} \, g_{\rho}^2 \, g_{\omega}^2 \, \rho_{03}^2 \,\, \omega_0 \;, 
\nonumber \\ 
m_\rho ^2 \, \rho_{03} &= \sum\limits_j^{}  g_{\rho j}\, \tau_{3j}\, n_{j}^{} 
+  \sum\limits_{b}^{} g_{\rho b} \, \tau_{3b} \, n_b
  \quad \nonumber \\
&- 2 \Lambda_{\rm v} \, g_{\rho}^2 \, g_{\omega}^2 \, \omega_0^2 \,\, \rho_{03} \;,
\label{eq-mov-rho03}
\end{align}
and
\begin{equation}
\zeta \left[ \mu_b - \omega_b^+(0) \right]\left[ \mu_b + \omega_b^-(0) \right] = 0   \; , 
\label{eq-mov-zeta}
\end{equation}
$ $
where 
\begin{equation}
n_{j}^s = \int \frac{d^3 p}{(2 \pi)^3} \frac{{m_j^*}}{E_j^*}  (f_{F+} + f_{F-})  \;,
\end{equation}
$ $
is the baryon scalar density of particle $j$, and the respective baryon density
\begin{equation}
 n_{j} =  \frac{2}{(2 \pi)^3} \int\limits_{}^{}{d^3 p \, (f_{F+} - f_{F-}) } \; .
\end{equation}

The ``boson scalar density'' for the boson $b$ is given by
\begin{equation}
n_b^s = \int_{}^{} {\frac{d^3 p}{(2 \pi)^3} \frac{m_b^*}{\epsilon_b^*} (f_{B+} + f_{B-} ) }   \; , 
\end{equation}
$ $
and the boson density is given by (\ref{boson-dens}). %

\myindent From the last equation of (\ref{eq-mov-zeta}) we obtain the conditions for the possibility 
of a boson condensate ($\zeta = 0$, no condensate), resulting
\begin{equation}
\mu_b = \omega_b^+(p=0) \quad \;\;\; {\rm or} \;\;\; \quad  \mu_b = - \omega_b^-(p=0) \;,
\label{mub-onset}
\end{equation}
depending on the signal of $\mu_b$ (either positive or negative). Thus
\begin{equation}
\mu_b = m_b^* \;+\; X_0 \quad \;\;\; {\rm or} \;\;\; \quad  \mu_b = - (m_b^* - X_0) \,,
\end{equation}
and
\begin{equation}
\mu_b - X_0 = m_b^*  \quad \;\;\; {\rm or} \;\;\; \quad  \mu_b - X_0 = - m_b^*  \,,
\end{equation}
such that the condition for the onset of the condensate state is
\begin{equation}
\nu_b \to m_b^*  \quad \;\;\; {\rm or} \;\;\; \quad  \nu_b \to - m_b^* \;\;.
\label{onset-cond}
\end{equation}
$ $

According to (\ref{boson-press}) and (\ref{onset-cond}) the condensate (zero 
momentum state) does not contribute to the pressure of the system as expected. 
When the condensate is not present, $\zeta = 0$. 

%%%%%%%%%%%%%%%%%%%%%%%%%%%%%%%

\section{The Mixed Phase}
%\begin{center}

In the following, three situations for the phase coexistence
are discussed in detail: A) hadron matter constituted of nucleons and 
quark matter constituted of quarks u and d, B) hadron matter constituted of 
nucleons and pions and quark matter constituted of quarks u and d and
C)  hadron matter constituted of nucleons, hyperons, pions and kaons with
zero net strangeness and quark matter constituted of quarks u and d. 

%\medskip
\subsection{A. Nucleons and quarks}
%\medskip
%\end{center}

According to the Gibbs' conditions \cite{landau-gibbs} for the phase 
coexistence the chemical potentials, temperatures 
and pressures have to be identical in both phases (H = hadron phase; 
Q = quark phase):

\begin{equation}
\begin{array}{c}
\mu^{H}_u = \mu^{Q}_u   \;,  \\ 
\mu^{H}_d = \mu^{Q}_d   \;,  \\
T^{H}=T^{Q}   \;,  \\ 
P^{H}(\mu^{H}_u,\mu^{H}_d,T)=P^{Q}(\mu^{Q}_u,\mu^{Q}_d,T)  \;.  \\ 
\end{array}
\label{gibbs1}
\end{equation}
$ $

The conservation of the isospin ($n_3$) and baryon densities ($n_B$)
are also required, so that in terms of these two charges \cite{muller-serot} 
and including the mixed phase we can write
\begin{equation}
\begin{array}{c}
P^{H}(n_B^H, n_3^H, T) = P^{Q}(n_B^Q, n_3^Q, T)   \;,  \\ \\
\mu_B^{H}(n_B^H, n_3^H, T) = \mu_B^{Q}(n_B^Q, n_3^Q, T)   \;,  \\
\mu_3^{H}(n_B^H, n_3^H, T) = \mu_3^{Q}(n_B^Q, n_3^Q, T)   \;,  \\  \\
n_B = (1 - \chi)n_B^H + \chi n_B^Q  \;,  \\ 
n_3 = (1 - \chi)n_3^H + \chi n_3^Q  \;,  \\ 
\end{array}
\end{equation}
%$ $

where for the hadron phase
\begin{equation}
\begin{array}{c}
\; n_B^H =  n_p + n_n  \quad \quad \quad ; \;\; \quad  n_3^H = \dfrac{n_p - n_n}{2}   \;,  \\ 
\\
\mu_B^H = \dfrac{1}{2}(\mu_p + \mu_n)  \quad \;\;\; ; \;\;\; \quad  \mu_3^H = \mu_p - \mu_n   \;,  \\
\end{array}
\end{equation}
%$ $

and for the quark phase

\begin{equation}
\begin{array}{c}
n_B^Q = \dfrac{1}{3}(n_u + n_d)  \quad \;\;\; ; \;\;\; \quad  n_3^Q = \dfrac{n_u - n_d}{2}   \;,  \\ 
\\
\mu_B^Q = \dfrac{3}{2}(\mu_u + \mu_d)  \quad \;\;\; ; \;\;\; \quad  \mu_3^Q = \mu_u - \mu_d   \;,  \\
\end{array}
\end{equation}
$ $

and $\chi$ represents the fraction of quarks in the mixed phase. 
The asymmetry parameter, $\alpha$ (isospin ratio) of the nuclei was defined as

\begin{equation}
\alpha \equiv \, \frac{N-Z}{N+Z} = \frac{n_n-n_p}{n_B}   \;,
\end{equation}
$ $

and the asymmetry parameter of the hadron and quark phases can be defined by:
\begin{equation}
\alpha^H \equiv -2 \, \frac{n_3^H}{n_B^H} \quad ; \quad \alpha^Q \equiv -2 \, \frac{n_3^Q}{n_B^Q}  \;,
\end{equation}
hence
\begin{equation}
\alpha^H = \frac{n_n - n_p}{n_n + n_p} \quad ; \quad \alpha^Q = 3 \frac{n_d - n_u}{n_d + n_u} \;,
\end{equation}
$ $

such that $0 \leq \, \alpha^H \leq 1$ (just nucleons case) and the quark 
one $0 \, \leq \alpha^Q \leq 3$.

%\begin{center}
%\medskip
\subsection{Nucleons, pions and quarks}
%\medskip
%\end{center}

When bosons are present the isospin density 
of the hadron phase is modified according to (\ref{eq-mov-rho03}) and $\alpha^H$ 
can be greater than 1. The isospin density of the hadron phase with $\pi^-$ becomes
\begin{equation}
n_3^H = \dfrac{n_p - n_n}{2} - n_{\pi}  \;,
\end{equation}

where $n_{\pi} = n_{\pi}^c + n_{\pi}^T$ and we assume $g_{\rho N} = g_{\rho} = g_{\pi}$. 
In order to obtain the simplest
thermodynamic features for the pions we set $g_{\omega \pi} = 0$ and for 
simplicity $g_{\sigma \pi} = 0$ so that in this case $m_{\pi}^* = m_{\pi}$. 
The in-medium (s-wave) Bose effective pion energy is
\begin{equation}
\omega_{\pi^-}(p=0) = m_{\pi}- g_{\rho}~\rho_{03}  \;,
\label{omegapi}
\end{equation}
and the pion chemical potentials and the effective $\pi^-$ chemical potential
\begin{equation}
\mu_{\pi^-} = \mu_n - \mu_p \quad,\quad 
\mu_{\pi^+} = - \mu_{\pi^-} \quad,\quad 
\end{equation}
\begin{equation}
\mu_{\pi^0} = 0 \quad , \quad 
\nu_{\pi^-} = \mu_{\pi^-} + g_{\rho}~\rho_{03}  \;.
\end{equation}
%$ $

As $\mu_n > \mu_p$ then $\mu_{\pi^-} > 0$ and according to (\ref{onset-cond}) the onset of the pion ($\pi^-$)
condensation takes place when 
\begin{equation}
\nu_{\pi^-} \to m_{\pi}  \;.
\end{equation}
$ $

and the EoS for the hadronic phase

\begin{equation}
P_{H} = P_B + P_{\pi}  \quad ; \quad  {\cal E}_{H} = {\cal E}_B + {\cal E}_{\pi}  \;,
\end{equation}
$ $

where $P_{\pi}$ and ${\cal E}_{\pi}$ are given by (\ref{boson-press}) and 
(\ref{boson-energ}) and we have also included the neutral pions as a free Bose gas.

%%%%%%%%%%%%%%%%%%%%%%%%%%%%%%%%%%%%%%%
%\begin{center}
%\medskip
\subsection{The baryon octet, pions, kaons and quarks}
%\medskip
%\end{center}

\myindent  At this stage we have included in the hadron phase all baryons of the 
baryon octet and in order to keep the strangeness conservation as 
$\sum\nolimits_i {S_i = 0} $ in both phases we also have included 
the $K^+$ meson in the hadron phase. For the Gibbs' conditions (\ref{gibbs1}) 
we need to add: 
$\mu^{H}_s = \mu^{Q}_s$~, so that we can write the chemical potential as 
$\mu_i = B_i \mu_B + I_{3i} \mu_3 + S_i \mu_S$, where $B_i, I_{3i}$ and $S_i$ are 
the baryonic, isospin and strangeness quantum numbers of particle $i$.

\myindent The equations for the baryons and bosons are already 
presented in this work. For the kaons we set $g_{\rho K} = g_{\rho}$, $g_{\omega K} = 0$ 
and also $g_{\sigma K} = 0$ such that $m_K^* = m_K$ as for the pions.
We are aware that this choice is very naive. It was done in order to  
explore the isospin degree of freedom. In a future work a more realistic 
parametrization of the kaon-meson coupling will be used which will allow us 
to discuss the strangeness degree of freedom more completely.

%%%%%%%%%%%%%%%%%%%%%%%%%%%%%%%%%%%%%%%%%%%%%%%%%%%%%%%%%%%%%%%
%
%%%%%%%%%%%%%%%%%%%%%%%%%%%%%%%%%%%%%%%%%%%%%%%%%%%%%%%%%%%%%%%
\section{Results}

\begin{figure}[!ht]
  \centering
\begin{tabular}{c}
\includegraphics[width=8cm,height=6.4cm,angle=0]{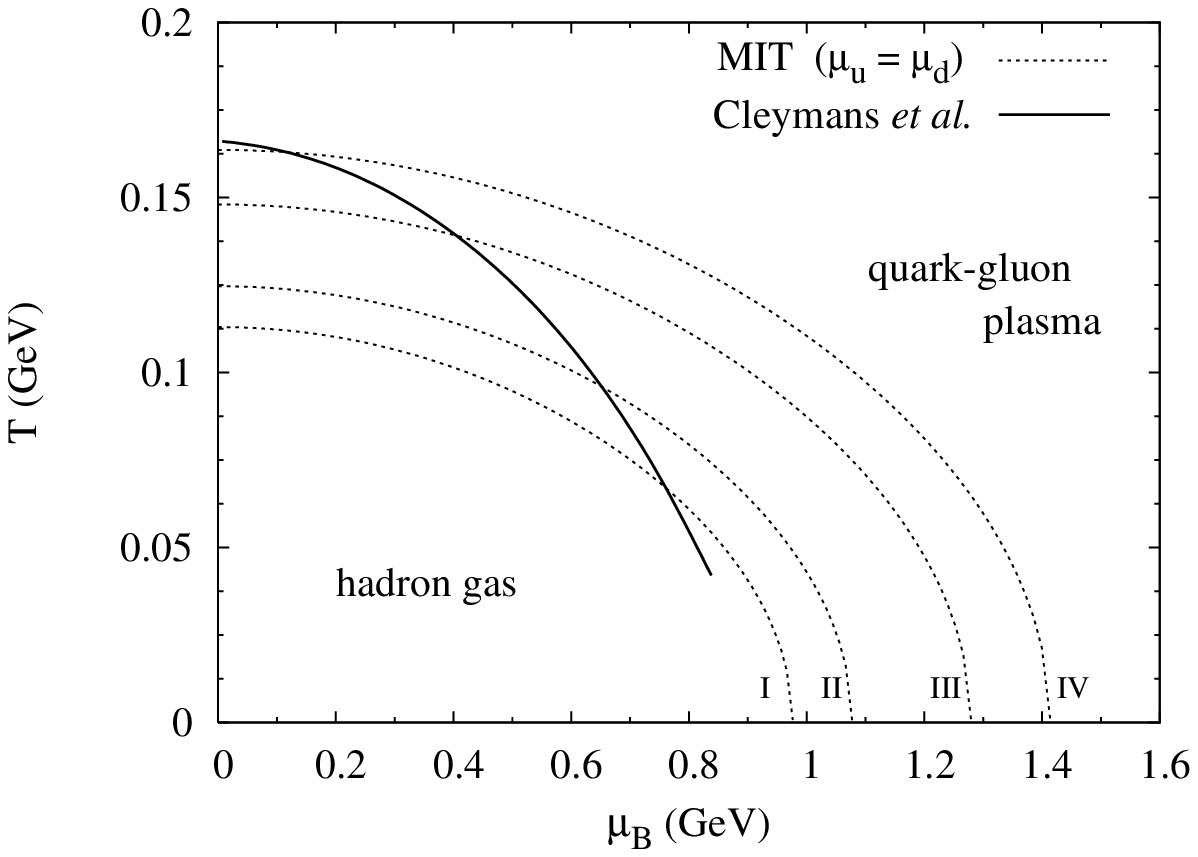} \\
 {\bf (a)} \\
\includegraphics[width=7.3cm,height=6.0cm,angle=0]{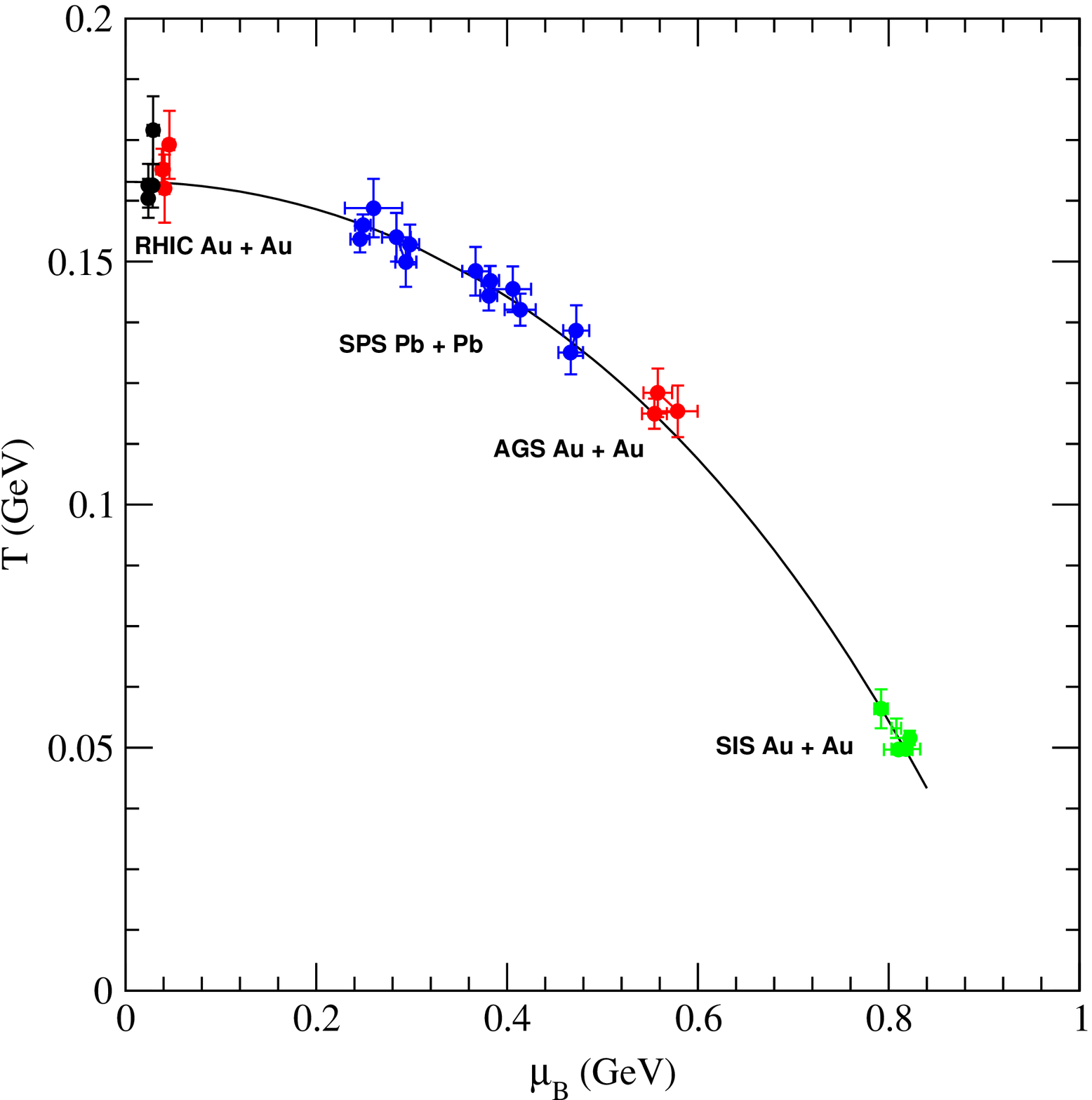} 
\\ {\bf (b)}
\end{tabular}
\caption{{\bf (a)} A simple qualitative overview on the T $\times$ $\mu_B$ curve in 
the MIT bag model (dashed curves) for the case $\mu_u = \mu_d$~, compared with the 
freeze-out curve (continuous line) from figure (b). The bag constant values from I 
to IV are: $B^{1/4}$ = 145, 160, 190 and 210 (MeV). {\bf (b)} A parametrization of 
the freeze-out curve deduced from particle multiplicities in heavy-ion collisions, 
Cleymans {\it et al.} (2006) \cite{t-mu-cleymans1}.}
 \label{mit-cley}
\end{figure}

\myindent First of all it is important to present some features of the MIT bag 
model. Figure \ref{mit-cley} shows a qualitative overview of the MIT bag model in a 
simple case when $\mu_u = \mu_d$~, comparing with 
data analysis of some collision experiments \cite{t-mu-cleymans1}. This figure also indicates the best 
values of the bag constant $B$ to be used in some energy ranges 
that we describe in the present work. In high 
temperatures and low baryon chemical potential (low density) we use the values 
$B^{1/4} = 190$ and 210 MeV. On the other hand, our analysis at intermediate energies is 
 performed with $B^{1/4} = 160$~MeV.

\begin{figure*}[ht]
  \centering
\begin{tabular}{cc}
\includegraphics[width=8cm,height=6.0cm,angle=0]{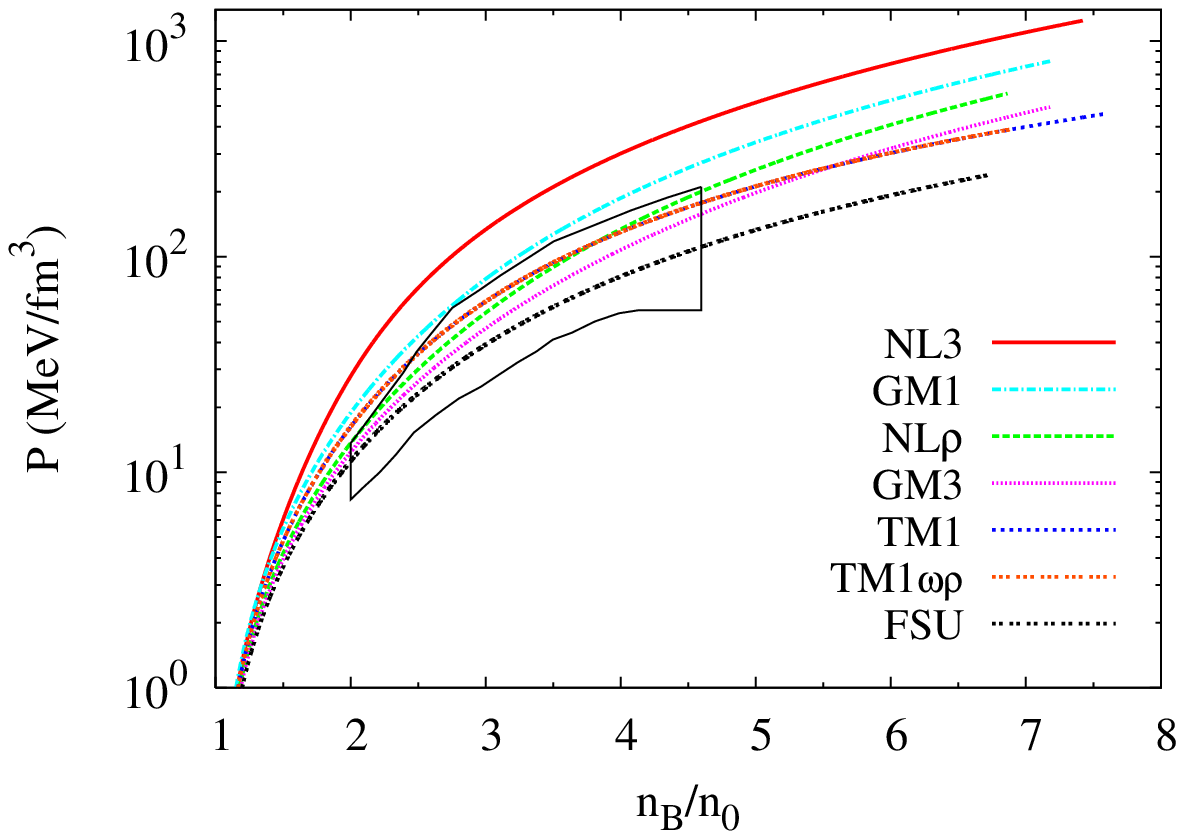}
 &
\includegraphics[width=8cm,height=6.0cm,angle=0]{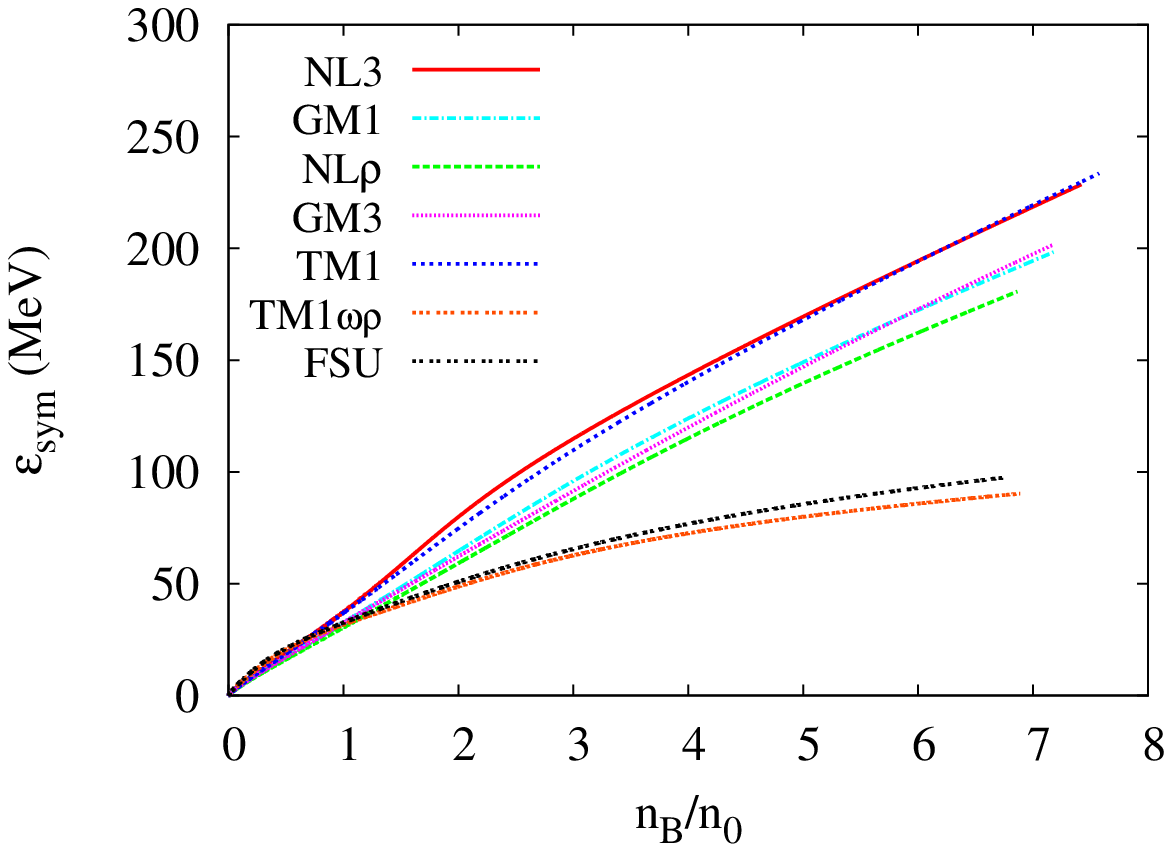}
\\ {\bf (a)} & {\bf (b)}
\end{tabular}
\caption{EoS for symmetric matter and different models. {\bf (a)} Pressure as a function of the baryon number 
density. The enclosed area represents experimental data according to \cite{danielewicz1}. 
{\bf (b)} The symmetry energy as a function of the baryon number density. }
 \label{press-ener-sym}
\end{figure*}

For the hadronic phase we use the parameter sets presented in table
\ref{tab-parameters}, where we give the symmetric nuclear matter properties 
at saturation density as well as the parameters of the models. 
In Figs.  ~\ref{press-ener-sym} (a) and (b)  the pressure of symmetric nuclear matter and
the symmetry energy, respectively, are  plotted for a large range of densities. In  Fig.
~\ref{press-ener-sym} (a) we also include the experimental constraints   obtained from
collective flow data in
heavy-ion collisions  \cite{danielewicz1}. We have considered a wide range of models frequently
used to study stellar matter or finite nuclei. Even though some of them 
do not satisfy the  constraints
determined in  \cite{danielewicz1}, as a whole these sets of models allows us 
to understand
the influence of a hard/soft equation of state (EOS) and a hard/soft symmetry energy of the
hadron matter-quark matter phase transition.  We have considered: 
 NL3 \cite{nl1-3-lalazissis1} with a quite large symmetry energy and
incompressibility at saturation and which was fitted in order to
reproduce the ground state properties of both stable and unstable
nuclei, TM1 \cite{tm1-sumiyoshi1}  which also reproduces the ground
state   properties of both stable and unstable nuclei and provides
an equation of state of nuclear matter similar to the one obtained
in the RBHF (Relativistic Brueckner Hartree-Fock) theory, softer
than NL3 at high densities,   TM1$\omega\rho$ \cite{tm1-helena}, the TM1 parametrization with a
mixed isoscalar-isovector coupling which we fix in order to
obtain a softer  density dependence of the symmetry energy
\cite{piekarewicz1}, FSU \cite{fsu} which was
accurately calibrated to simultaneously describe the GMR in
$^{90}$Zr and $^{208}$Pb, and the IVGDR in $^{208}$Pb and still
reproduce ground-state observables of stable and unstable nuclei;
GM1  and GM3  \cite{gm1-3-glendenning1} generally used to describe stellar matter, with a
symmetry energy not so hard as the one of NL3 and TM1, and 
 NL$\rho$  \cite{nlrho-liu}, which has been used to discuss the hadron matter-quark matter
transition in \cite{ditoro2010}, and which presents an EOS at high densities between GM1 and GM3.

\myindent Let us first describe some hadron-quark matter systems at zero and finite temperature, including the  deconfined 
phase transition,  through isothermal processes.
We first discuss the effects of pions and
gluons on the phase transition. For this discussion we take NL3 to describe the hadronic matter, however the main conclusions do not depend on the nuclear model considered.

\begin{table*}[ht]
\centering
\begin{tabular}{cccccccc}
\hline
 &  {\bf FSU} \cite{fsu}   &  {\bf TM1} \cite{tm1-sumiyoshi1}  &  {\bf TM1}$\omega\rho$ \cite{tm1-helena}  &  {\bf NL$\rho$} \cite{nlrho-liu}  &  {\bf NL3} \cite{nl1-3-lalazissis1}  &  {\bf GM1} \cite{gm1-3-glendenning1}  &  {\bf GM3} \cite{gm1-3-glendenning1} \\ 
\hline
$n_0$ (fm$^{-3}$) & 0.148  & 0.145 & 0.145 & 0.160 & 0.148 & 0.153 & 0.153 \\ 
$K$ (MeV) & 230  & 281 & 281 & 240 & 271.76 & 300 & 240 \\ 
$m^*/m$ & 0.62  & 0.643 & 0.643 & 0.75 & 0.60 & 0.70 & 0.78 \\ 
$m$~(MeV) & 939  & 938 & 938 & 939 & 939 & 938 & 938 \\ 
-$B/A$ (MeV) & 16.3  & 16.3 & 16.3 & 16.0 & 16.299 & 16.3 & 16.3 \\ 
${\cal E}_{\rm sym}$ (MeV) & 32.6  & 36.9 & 31.9 & 30.5 & 37.4 & 32.5 & 32.5 \\ 
$L$ (MeV) & 61  & 110 & 55 & 85 & 118 & 94 & 90 \\
\hline
$m_{\sigma}$ (MeV) & 491.5  & 511.198 & 511.198 & 512 & 508.194 & 512 & 512 \\ 
$m_{\omega}$ (MeV) & 782.5  & 783 & 783 & 783 & 783 & 783 & 783 \\ 
$m_{\rho}$ (MeV) & 763  & 770 & 770 & 763 & 763 & 770 & 770 \\ 
$g_{\sigma}$ & 10.592  & 10.029 & 10.029 & 8.340 & 10.217 & 8.910 & 8.175 \\ 
$g_{\omega}$ & 14.302  & 12.614 & 12.614 & 9.238 & 12.868 & 10.610 & 8.712 \\ 
$g_{\rho}$ & 11.767  & 9.264 & 11.147 & 7.538 & 8.948 & 8.196 & 8.259 \\ 
$b$ & \;\; 0.000756 \;\;  & \;\; -0.001506 \;\; & \;\; -0.001506 \;\; & \;\; 0.006935 \;\; &\;\; 0.002052 \;\; & \;\; 0.002947 \;\; & \;\; 0.008659  \\ 
$c$ & 0.003960  & 0.000061 & 0.000061 & -0.004800 & -0.002651 & -0.001070 & -0.002421 \\ 
$\xi$ & 0.06  & 0.0169 & 0.0169 & 0 & 0 & 0 & 0 \\ 
$\Lambda_{\rm v}$ & 0.03  & 0 & 0.03 & 0 & 0 & 0 & 0 \\ 
\hline
\end{tabular}
\caption{Parameter sets used in this work and corresponding saturation 
properties.}
\label{tab-parameters}
\end{table*}

\begin{figure*}[ht]
  \centering
\begin{tabular}{cc}
\includegraphics[width=8cm,height=6.2cm,angle=0]{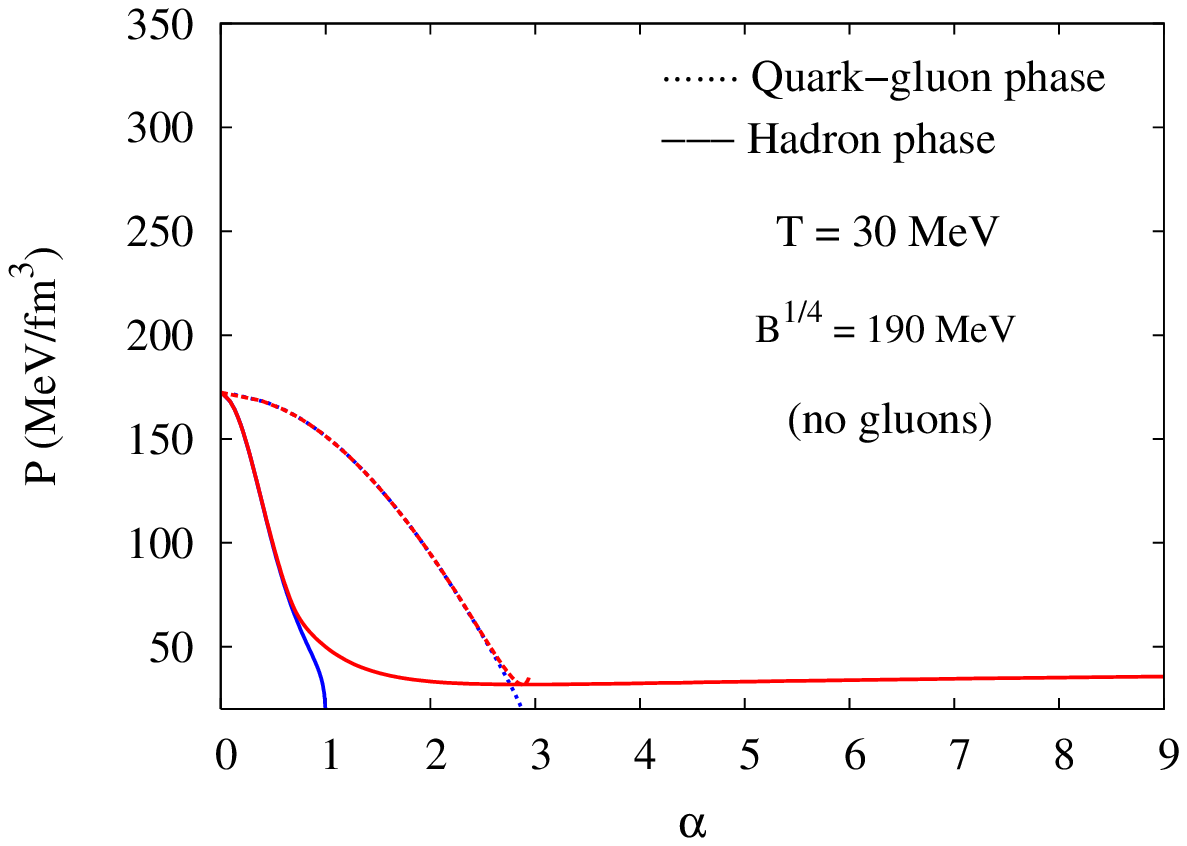}
 &
\includegraphics[width=8cm,height=6.2cm,angle=0]{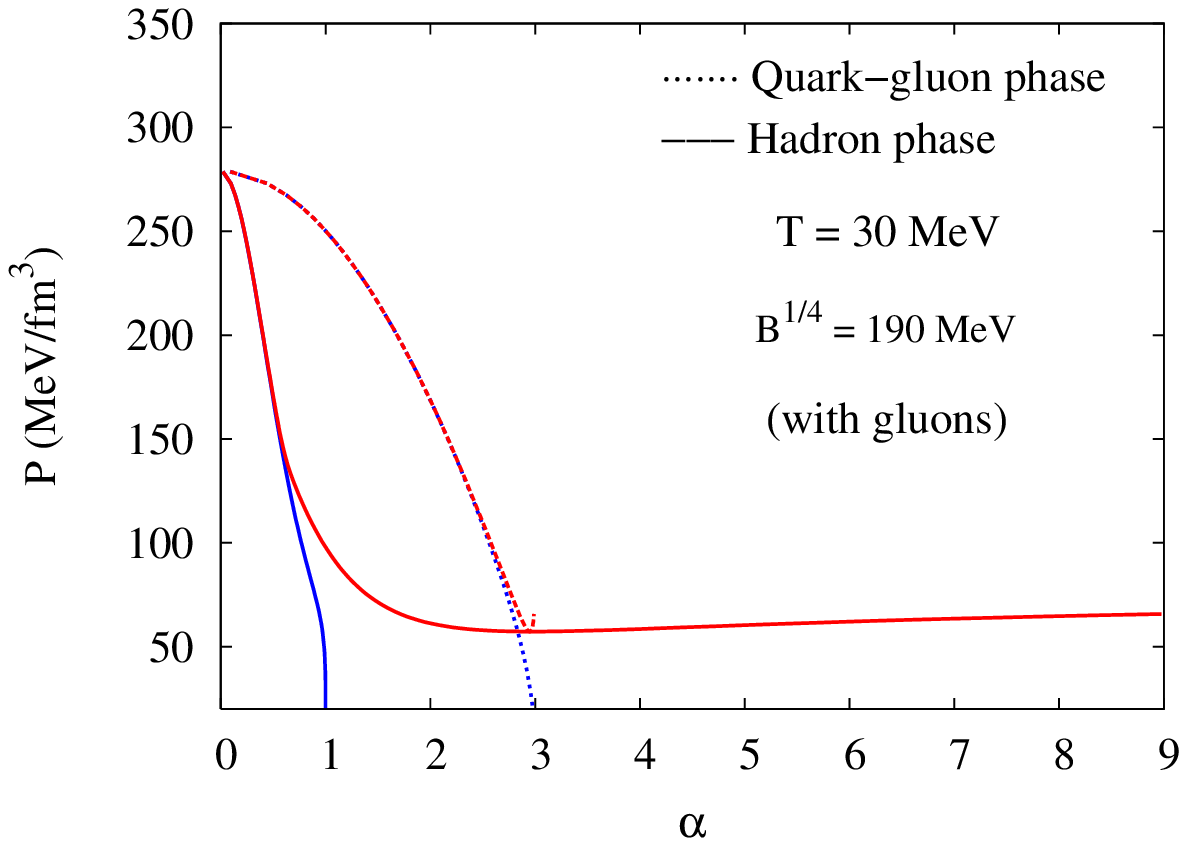} 
\\ {\bf (a)} & {\bf (b)}
\end{tabular}
\caption{Binodal sections at $T=30$~MeV and the effect of pions and gluons. The 
NL3 parameter set is used. The blue lines indicate a system with no pions. The red ones 
indicate the presence of pions. {\bf (a)} Without gluons. {\bf (b)} With gluons.}
 \label{press-assym-t30}
\end{figure*}

\myindent In Figs.~\ref{press-assym-t30} (a) and (b) we show slices of the binodal surface 
indicating the two-dimensional phase-coexistence boundary in \{$P, T, \alpha$\} space, 
at $T = 30$~MeV. For each temperature, the binodal section is divided into two branches. 
One branch describes the system in the hadron phase, while the other
branch describes the quark-gluon phase. 
In both figures the role of the pions and gluons is presented. Fig.~\ref{press-assym-t30} (a) 
shows a system with no gluons for two cases with and without the pions. 
The blue curves represent a system with no pions. 
The gluons have a very strong effect on the critical point, 
corresponding to a maximum in the pressure, when both phases coexist for 
$\alpha=0$. The presence of gluons increases the critical pressure almost by 
100\%.
The role of the pions is better seen in Fig.~\ref{press-assym-t30} (b) (red lines). 
The asymmetry parameter of the hadron phase 
increases due to the presence of the pions which increase the isospin interaction.
In equilibrium, the pressures in both phases must be equal according to the Gibbs' 
conditions. When pions are present these conditions still hold. We observe 
a slight increase of the pressure of the quark-gluon phase for $\alpha > 3$, following 
the hadronic pressure increase. 

\begin{figure}[ht]
  \centering
\includegraphics[width=8.6cm,angle=0]{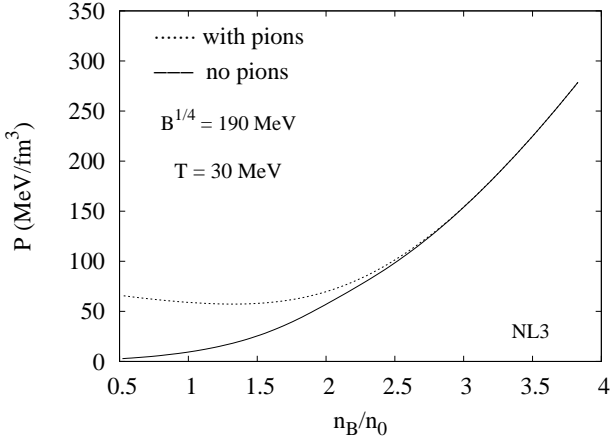}
\caption{Binodal section: pressure as a function of the baryonic density. The presence of the pions 
increases the isospin interaction such that a pressure increase is observed.}
 \label{press-rho-t30}
\end{figure}

\myindent At finite temperature pions are present as a Bose gas and  
 their presence as a condensate state at low enough temperatures is also
possible. The presence of a pion gas and a pion condensate changes the pressure at low densities 
according to Fig.~\ref{press-rho-t30} by increasing the 
absolute value of the $\rho$ meson field (i.e., the isospin interaction) since 
the condensate itself does not contribute to the pressure of the system as 
a boson gas. The lowest pressures of the binodal  occur for the largest values 
of the asymmetry parameter $\alpha$ (1 for the hadronic phase without pions).

\begin{figure}[ht]
  \centering
\includegraphics[width=8.6cm,height=6.2cm,angle=0]{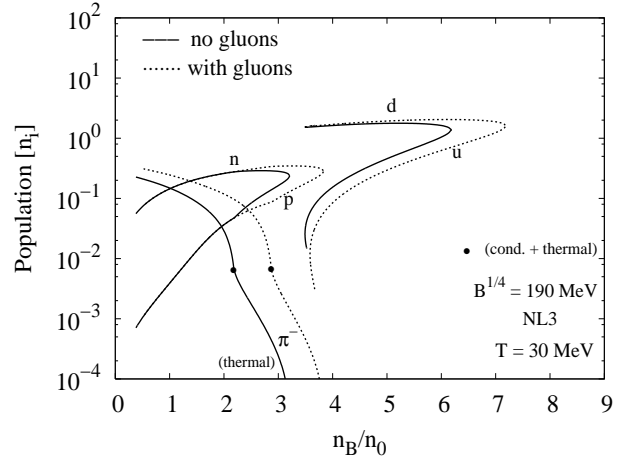}
\caption{Population of particles at $T = 30$~MeV with and without the gluons. The 
filled circles 
mark the onset of the pion condensate at $2.18 n_0$ in the first case and at 
$2.87 n_0$ in the second one.}
 \label{population0}
\end{figure}

\myindent When gluons are included in the quark phase the densities reached by the 
system at the binodal surface increase slightly in both 
phases such that the onset of the pion condensation takes place at a slightly higher 
density: $2.87 n_0$ instead of $2.18 n_0$ at $T = 30$~MeV according to 
Fig.~\ref{population0}, when an isothermal  
process is analyzed. Therefore, the presence of gluons shifts the phase transition 
to a quark-gluon plasma to larger densities.

\begin{figure}[hbt]
  \centering
\includegraphics[width=8.6cm,height=6.2cm,angle=0]{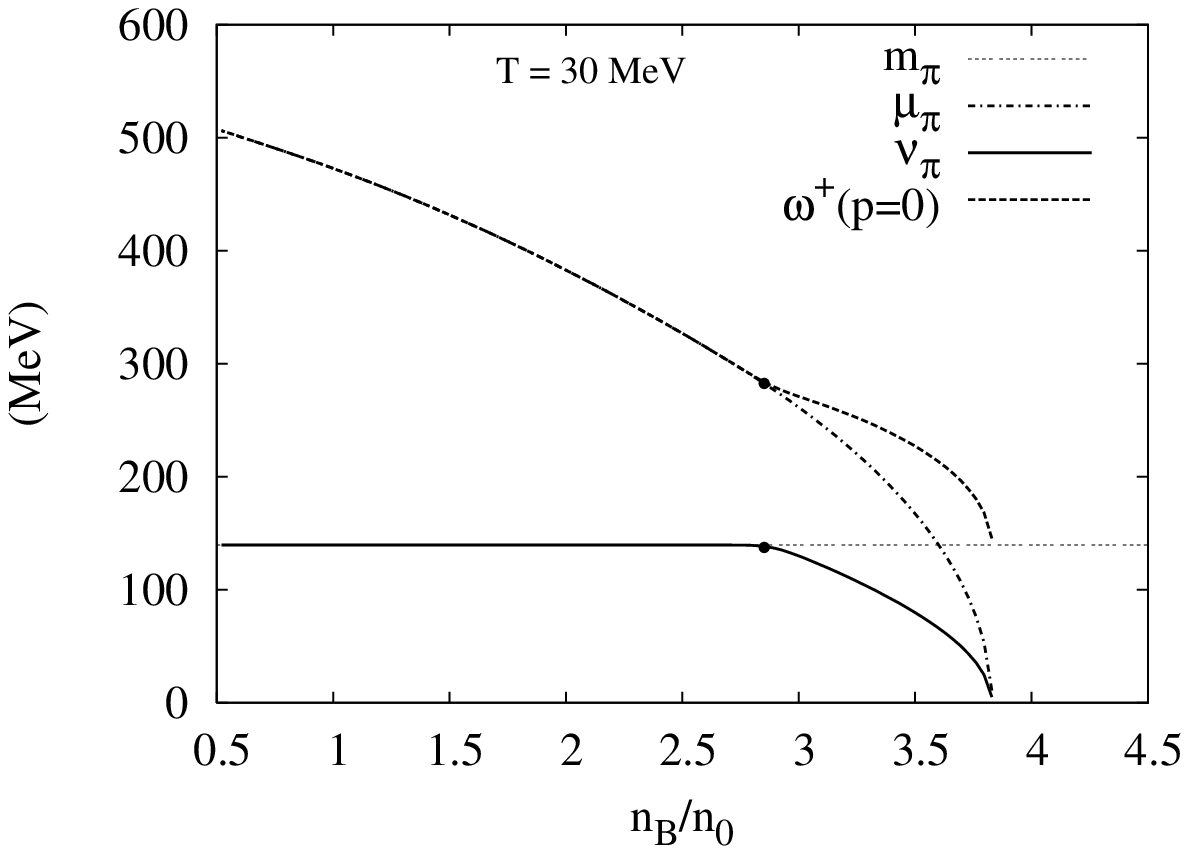}
\caption{The onset of the pion condensation for the case with gluons in
Fig.~\ref{population0}. The $y$-axis we plot the pion mass, chemical potential, 
effective chemical potential and frequency at $p=0$. }
 \label{press-assym-geral}
\end{figure}

\myindent The onset of the pion condensation according to the eq. (\ref{onset-cond}) 
[or similarly eq. (\ref{mub-onset})] is clearly seen  in Fig.~\ref{press-assym-geral} 
where we plot the pion mass $m_\pi$, pion chemical potential $\mu_\pi$, pion 
effective chemical potential $\nu_\pi$ and pion frequency at $p=0$, $\omega^+_b(p=0)$. 
The pion condensation occurs for the lower densities when the conditions  
(\ref{onset-cond}) or (\ref{mub-onset}) are satisfied.

\begin{figure}[ht]
  \centering
\includegraphics[width=8.6cm,height=6.0cm,angle=0]{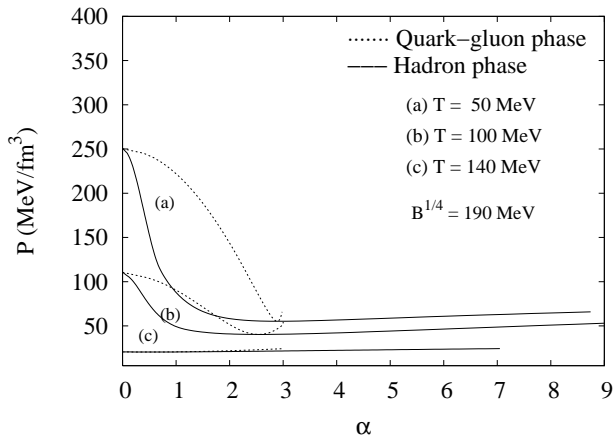}
\caption{Slices of the binodal surface using the NL3 parameter set and 
$B^{1/4} = 190$~MeV. The critical temperature is $T_c \sim 150$~MeV.  
The calculation includes both pions and gluons.} 
 \label{press-assym2}
\end{figure}

%%%%%%%%%%%%%%%%%%%%%%%%%%%%%%%%%%%%%%%%%%%%%%%%%%%%%%%%%%%%%
\begin{figure*}[ht]
  \centering
\begin{tabular}{cc}
\includegraphics[width=8cm,height=6.2cm,angle=0]{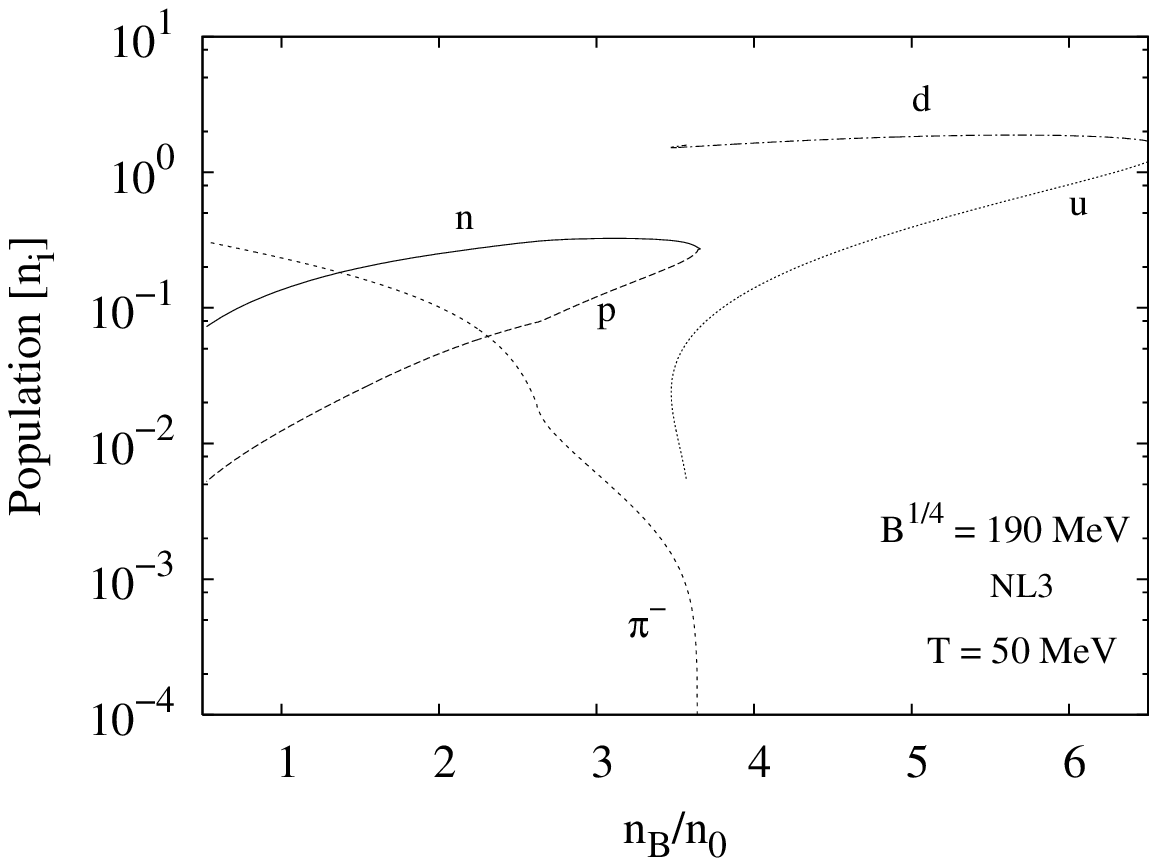}
&
\includegraphics[width=8cm,height=6.2cm,angle=0]{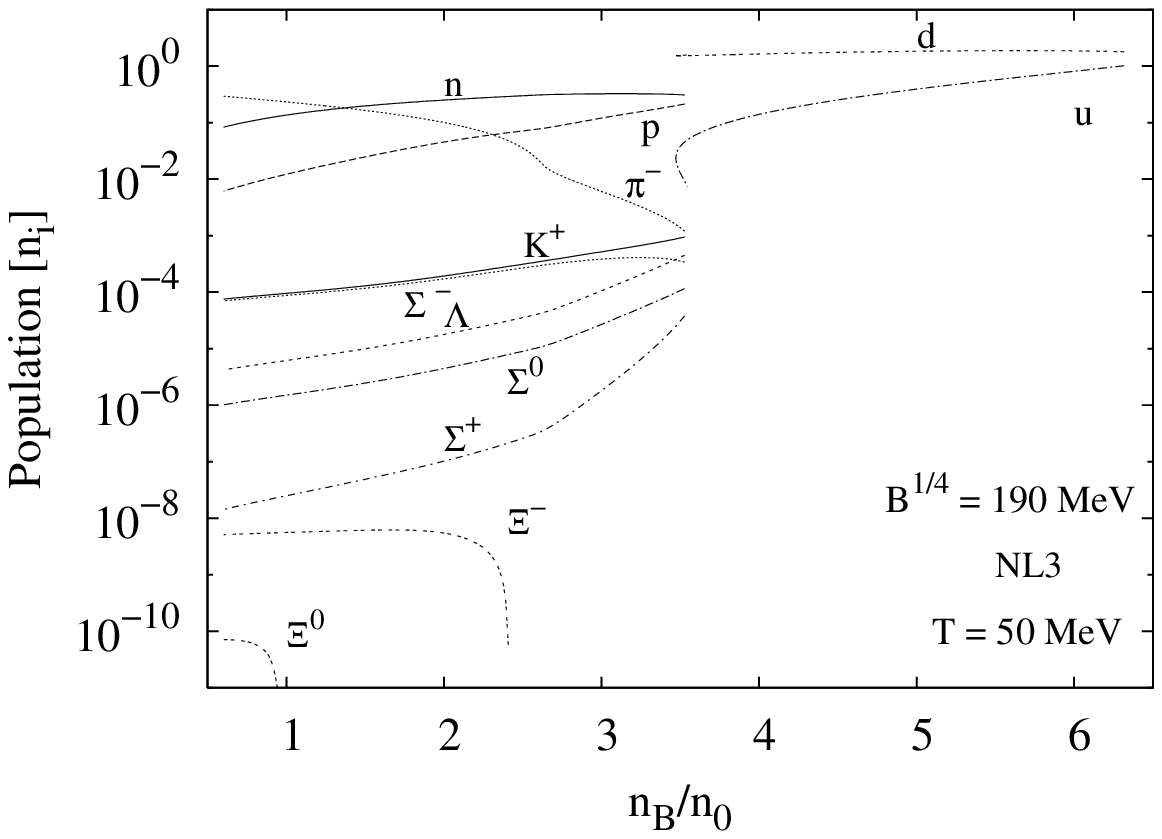}
\\ {\bf (a)} & {\bf (b)} 
\end{tabular}
\caption{Population of particles for the case in Fig.~\ref{press-assym2} (a) at 
$T = 50$~MeV. {\bf (a)} Just p, n and 
$\pi^-$ in the hadronic phase. {\bf (b)} The baryon octet, $\pi^-$ and $K^+$ in the 
hadronic phase.}
 \label{fig-population}
\end{figure*}
%%%%%%%%%%%%%%%%%%%%%%%%%%%%%%%%%%%%%%%%%%%%%%%%%%%%%%%%%%%%%
%
In Fig.~\ref{press-assym2}  we show the binodal slices 
at different temperatures and for  the bag constant 190~MeV. 
The enclosed area becomes smaller with increasing temperature and the pressure 
at $\alpha = 0$ decreases when the temperature increases. 
The two branches merge into a single line when the system reaches the 
critical temperature at zero chemical potential and density. The critical 
temperature ($T_c$) of the phase transition is $\sim 150$~MeV for the 
bag constant $B^{1/4} = 190$~MeV. For larger values of $B$ we obtain  
a larger pressures at the same temperature and the other way round for smaller values.
The  results shown in the figure  are consistent with the ones found in ref. \cite{muller-pions} 
although here the NL3 parameter set has been used. The calculation includes both 
pions and gluons.

\myindent Next we discuss the inclusion of strangeness. 
The population of particles at $T = 50$~MeV can be seen in 
Fig.~\ref{fig-population} for two cases: (a) a simple system of protons, 
p, neutrons, n, and pions, $\pi^-$, in the hadron phase, and (b) including 
the hyperons of the baryon octet and $K^+$ mesons in the hadron phase. 
In both cases the total strangeness of the system is zero, therefore,  we just have 
quarks u and d in the quark phase. Fig.~\ref{fig-population} (a) shows an 
increase of pions at low baryon densities, which plays an important role in 
the isospin density of the system. Most of the pions below $2.6 n_0$ 
are in a zero momentum state (i.e., a pion condensate). The same pattern can be 
seen in Fig.~\ref{fig-population} (b) on pions and nucleons, indicating that 
strange particles are not important in these conditions at that temperature 
but they do appear at 
higher densities. We do not see kaon condensation, just a pion condensate as 
in the first case.  It is important to analyse how sensitive are the above results
to the choice of the kaon-meson interaction. Work in this direction will be 
done in the near future.

\begin{figure*}[hbt]
  \centering
\begin{tabular}{cc}
\includegraphics[width=8cm,height=6.2cm,angle=0]{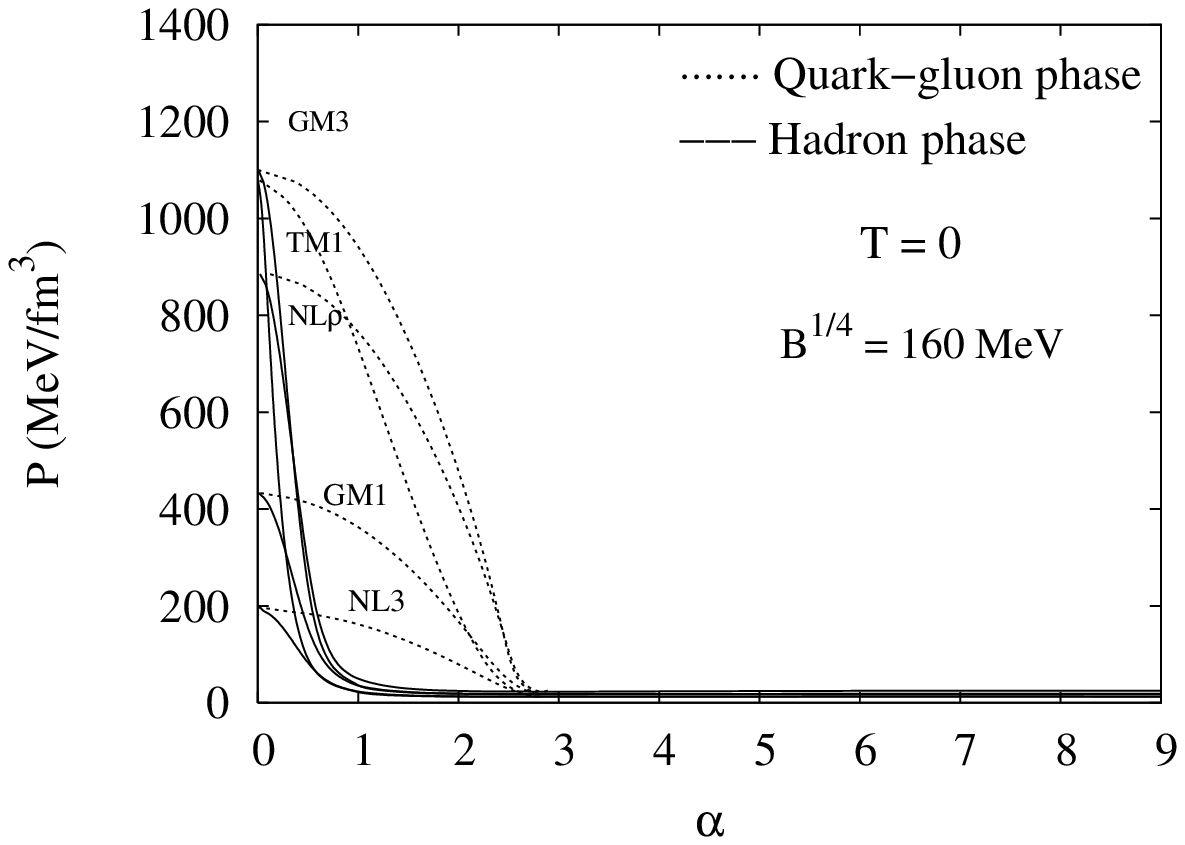}
 &
\includegraphics[width=8cm,height=6.2cm,angle=0]{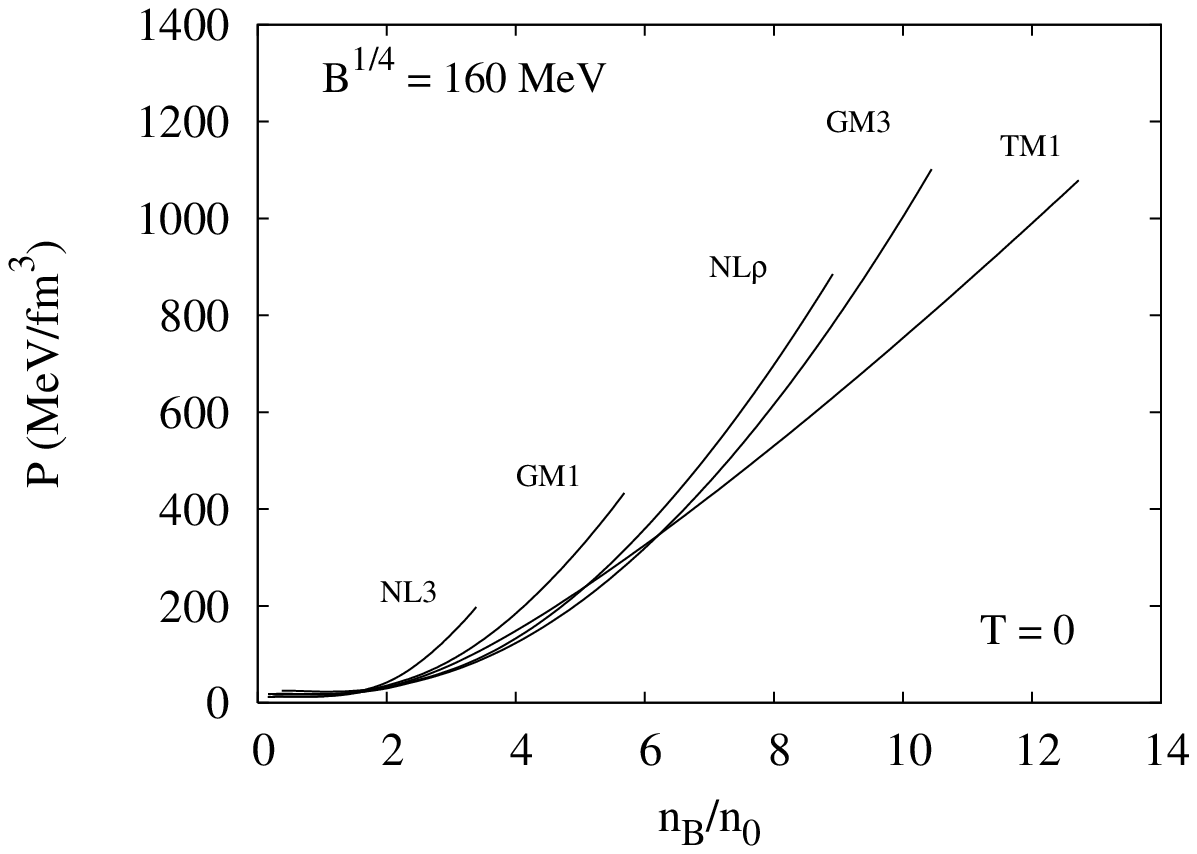}
\\ {\bf (a)} & {\bf (b)}
\end{tabular}
\caption{{\bf (a)} Pressure as a function of the asymmetry parameter for different 
parametrizations at zero temperature. {\bf (b)} Pressure as a function of the 
baryon number density for the case in figure (a).}
 \label{press-assym-t0-2}
\end{figure*}

\begin{figure}[ht]
\includegraphics[width=8.6cm,height=6.2cm,angle=0]{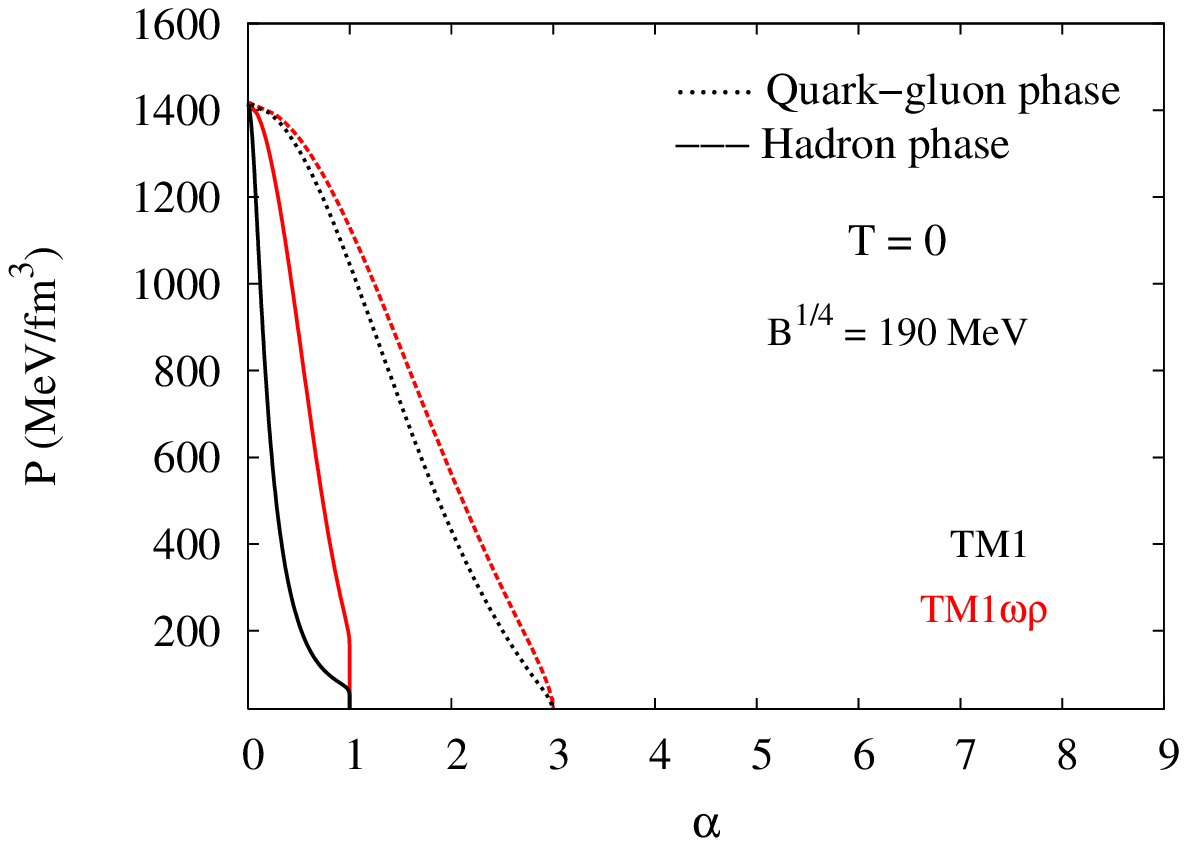} 
\caption{The effect of the $\omega-\rho$ coupling on the binodal at $T=0$.}
 \label{press-assym-t30-wr}
\end{figure}

\myindent We now discuss the effect of the density dependence of the EOS on the binodal 
surfaces.
In Fig.~\ref{press-assym-t0-2} (a) one sees a comparison of the hadron phase-quark 
phase binodal sections among the 
different parameter sets listed in Table \ref{tab-parameters} for the zero temperature 
case and $B^{1/4} = 160$~MeV. Qualitatively all the curves behave in the same way. 
The difference lies in the 
pressures and the densities reached by the different systems which is explicitly shown in 
Fig.~\ref{press-assym-t0-2} (b).
We conclude that the different behaviors seen for the binodal sections  are due
to the EoS at large densities, see Fig.~\ref{press-ener-sym} (a) where the 
pressure is
plotted as a function of density for cold symmetric nuclear matter. 
At finite temperature a
similar trend is obtained except that the maximum densities reached are smaller.

\myindent The effect of the symmetry energy on the binodal is better discussed analyzing
Fig.~\ref{press-assym-t30-wr} where the binodal for TM1 and TM1$\omega\rho$ is plotted without
pions. These two models have the same isoscalar behavior and just differ in the isovector
channel, TM1$\omega\rho$ having a softer symmetry energy. We conclude that  a softer symmetry
energy favors a phase transition at larger asymmetries. For the same reason the models with a
softer symmetry energy have their binodals for larger asymmetries in 
Fig.~\ref{press-assym-t0-2} (a).

\myindent Fig.~\ref{press-ener-sym} (a) and (b) are used in the following to discuss 
the differences between the models. We  see that the hadron density at the 
binodal surface is very
sensitive to the softness/hardness of the EoS at intermediate/high densities. In 
particular, the largest pressures are attained by the softest EOS. It is interesting 
to analyse the behavior of TM1:  it behaves  at low densities as a hard EOS like NL3 
and at high densities as a soft one, giving the largest pressure at the critical point. 
It is the relative change of hard/soft character of the EOS that explains  the 
crossing between the different models in Fig.~\ref{press-assym-t0-2} (a). We have 
not included a curve for FSU because due to its softness no phase transition was 
obtained at reasonable densities. The behavior at large densities can be adjusted 
by changing the value of the parameter $\chi$ which multiplies the forth power of 
the $\omega$-meson term  in the Lagrangian density. A larger value gives a softer 
EOS at large densities.  We have reduced the value of $\chi$ and for $\chi=0.03$ 
we could get convergence at reasonable densities. This coincides with the large 
density behavior of the new parametrization proposed in \cite{iu-fsu}, that corrects 
the behavior of FSU at large densities which predicted too small maximum star 
masses and too large star radii.

\myindent The density dependence of the
energy density does not affect the binodal surface of symmetric nuclear 
matter but it certainly has an effect if we consider asymmetric matter. We 
investigate the phase transition at intermediate energies 
using a  convenient choice of different parametrizations of the NLWM in order to explore
different compressibilities at large densities as well as an asysoft and asyhard EoS. We take into account the parameter sets: NL3, hard EoS and symmetry 
energy;  NL$\rho$, intermediate behavior both in
the isoscalar and isovector channel; TM1, soft EoS at high densities
and hard symmetry energy and TM1$\omega\rho$,  with a soft symmetry energy. 

\begin{figure}[ht]
  \centering
\begin{tabular}{c}
\includegraphics[width=8.6cm,height=6.2cm,angle=0]{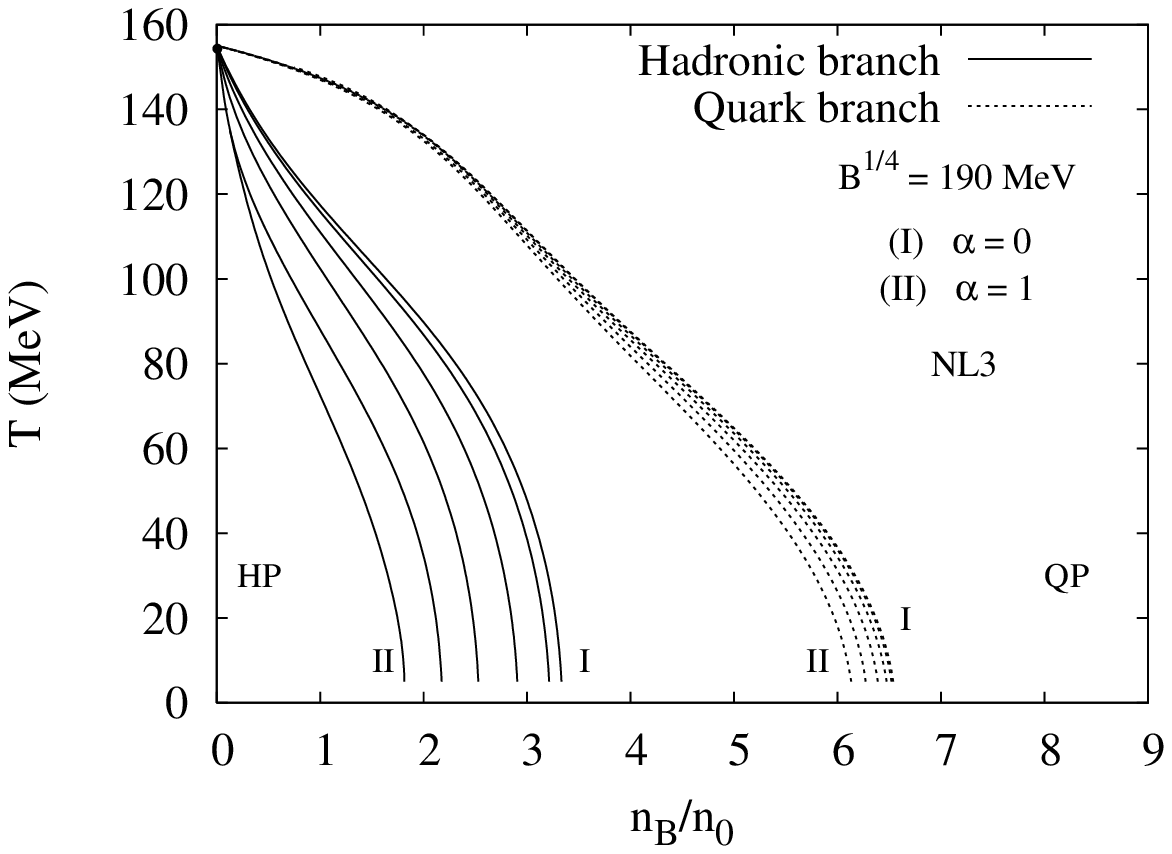} \\
 {\bf (a)} \\
\includegraphics[width=8.6cm,height=6.2cm,angle=0]{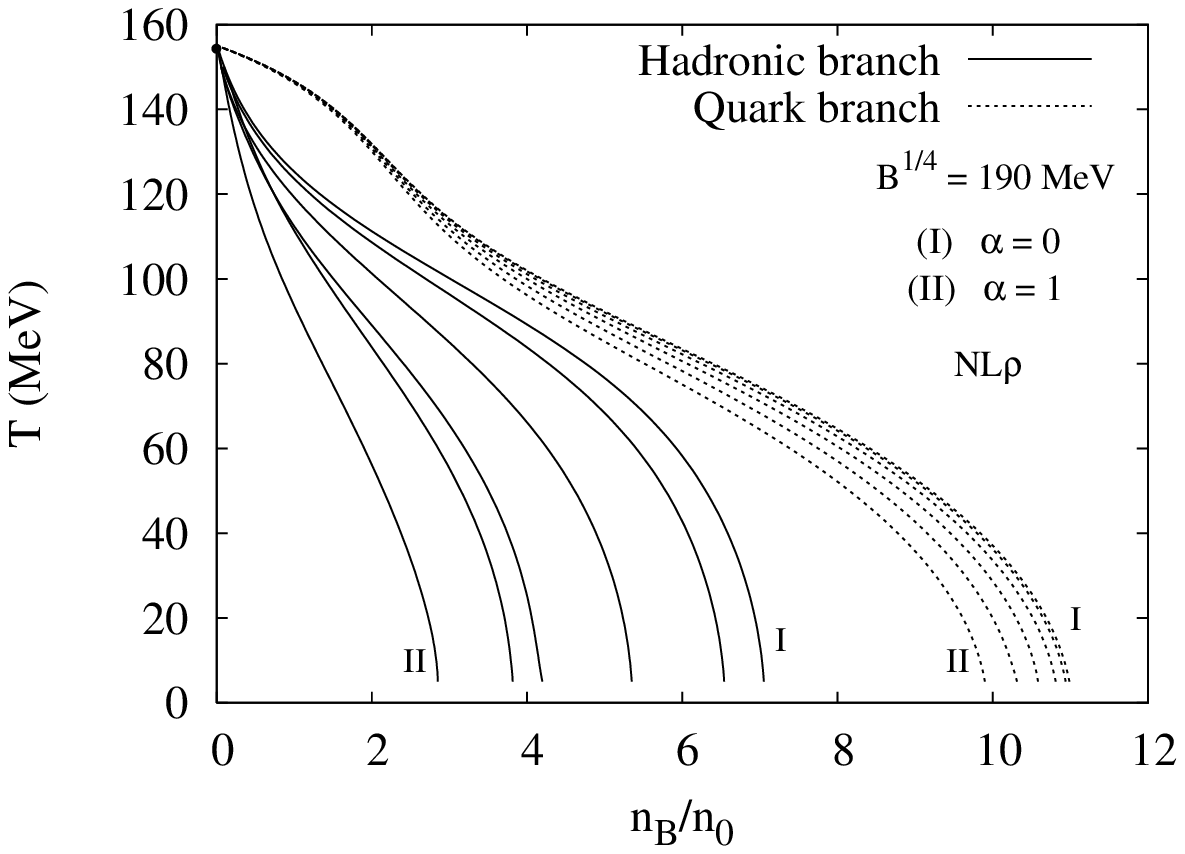} 
\\ {\bf (b)}
\end{tabular}
\caption{{\bf (a)} Binodal section in ($n_B$, $T$, $\alpha$) space and the projection 
of several branches for  different asymmetry parameter $\alpha$ onto the ($n_B$, $T$) plane. The 
asymmetries are: $\alpha = 0,~0.2,~0.4,~0.6,~0.8,~1.0$, from the right (I) to the left (II) in the 
two phases, with no gluons. {\bf (b)} Same as figure (a) for the NL$\rho$ parameter set. 
In both cases the critical temperature where $\mu_B = 0$ is $T_c \sim 155$~MeV.}
 \label{temp-densb}
\end{figure}

\myindent In order to discuss the effect of isospin asymmetry on the binodal
sections, we allow the temperature to change with fixed asymmetry 
parameter, and compare the predictions of the different models. Figs.~\ref{temp-densb} (a) and (b) show,  for NL3 and NL$\rho$, the binodal sections in 
\{$n_B, T, \alpha$\} space and the projection of several branches at 
different $\alpha$ onto the ($n_B, T$) plane (HP = hadron phase; QP = quark phase). 
In other words Figs.~\ref{temp-densb} 
(a) and (b) show the QCD phase diagram with different asymmetries, 
$\alpha = 0,~0.2,~0.4,~0.6,~0.8,$~and $1.0$, from the 
right (I) to the left (II) in the two phases. From now on, in order not to reach too 
high densities in the hadron phase we exclude the gluons from the system.
This does not affect the
comparison between models and may give rise to a maximum 20\% underestimation of the transition density.

\begin{figure*}[htb]
  \centering
\begin{tabular}{cc}
\includegraphics[width=8cm,height=6.2cm,angle=0]{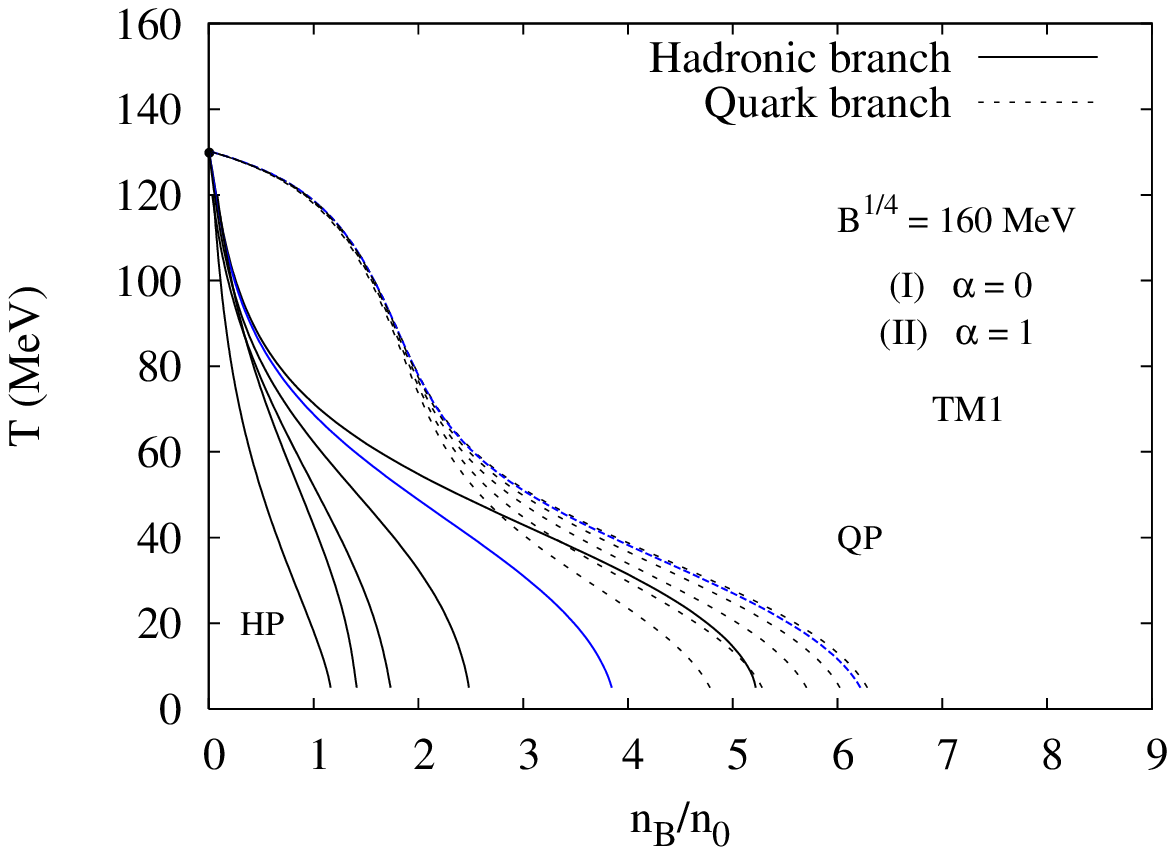}
 &
\includegraphics[width=8cm,height=6.2cm,angle=0]{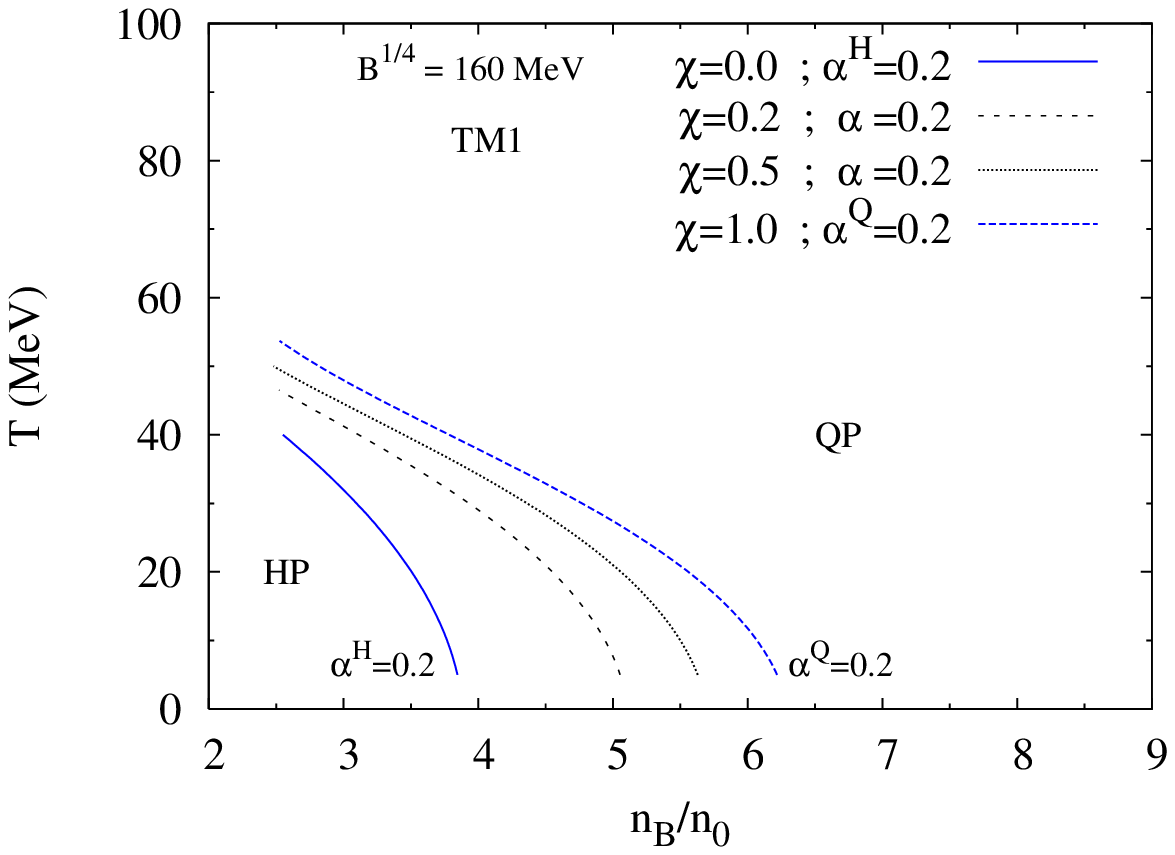}
\\ {\bf (a)} & {\bf (b)}

\\
\\

\includegraphics[width=8cm,height=6.2cm,angle=0]{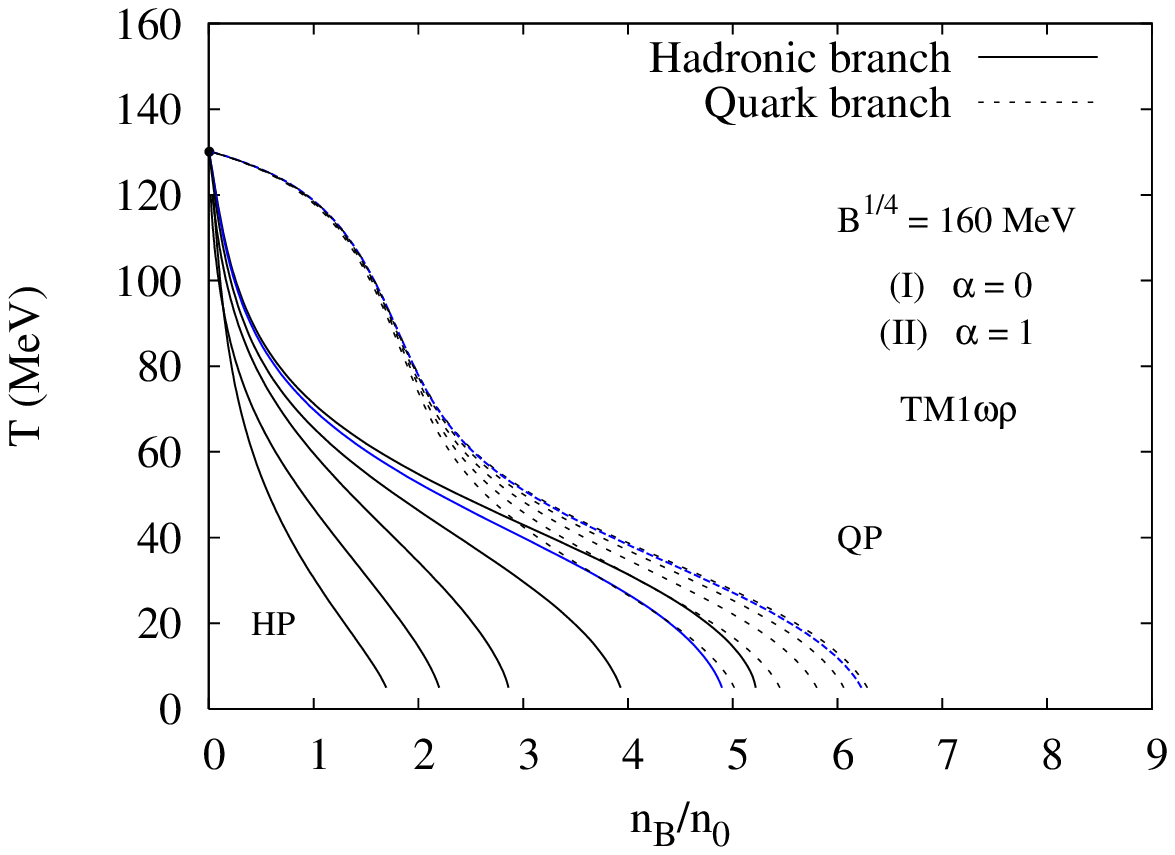}
 &
\includegraphics[width=8cm,height=6.2cm,angle=0]{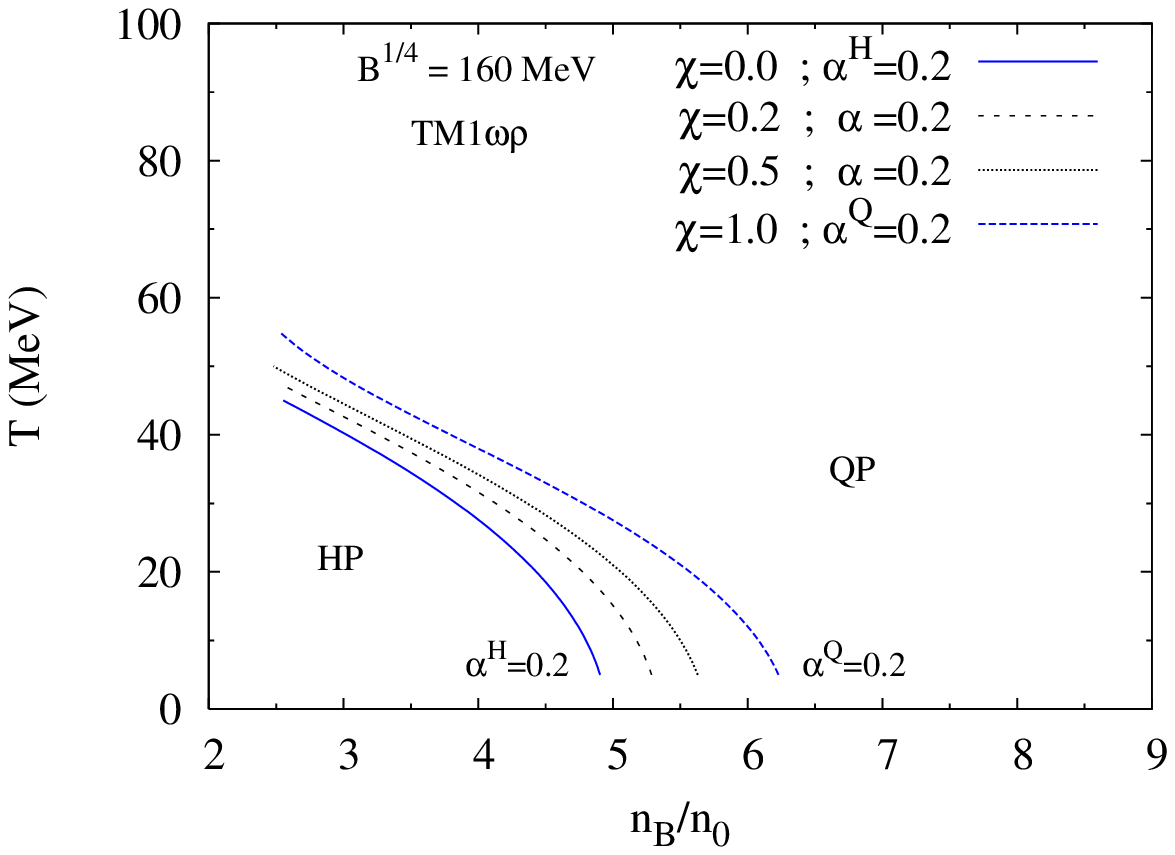}
\\ {\bf (c)} & {\bf (d)}

\end{tabular}

\caption{Same as in Fig.~\ref{temp-densb} for $B^{1/4} = 160$~MeV. The critical temperatures 
are $T_c \sim 130$~MeV and the labels (I) and (II) also represent the asymmetry 
in the same way as in Fig.~\ref{temp-densb}. {\bf (a)} The TM1 parameter set is used. {\bf (b)} 
Part of Fig.~\ref{press-assym-tm1} (a) for $\alpha = 0.2$ with the mixed phase for 
different quark concentrations ($\chi = 0.2, 0.5$).
{\bf (c)}  The TM1 parameter set and the mixing term $\Lambda_{\rm v}$ has been used in the 
present case. {\bf (d)} Part of Fig.~\ref{press-assym-tm1} (c) with the mixed phase. 
}
 \label{press-assym-tm1}
\end{figure*}
\begin{figure*}[htb]
  \centering
\begin{tabular}{cc}
\includegraphics[width=8cm,height=6.2cm,angle=0]{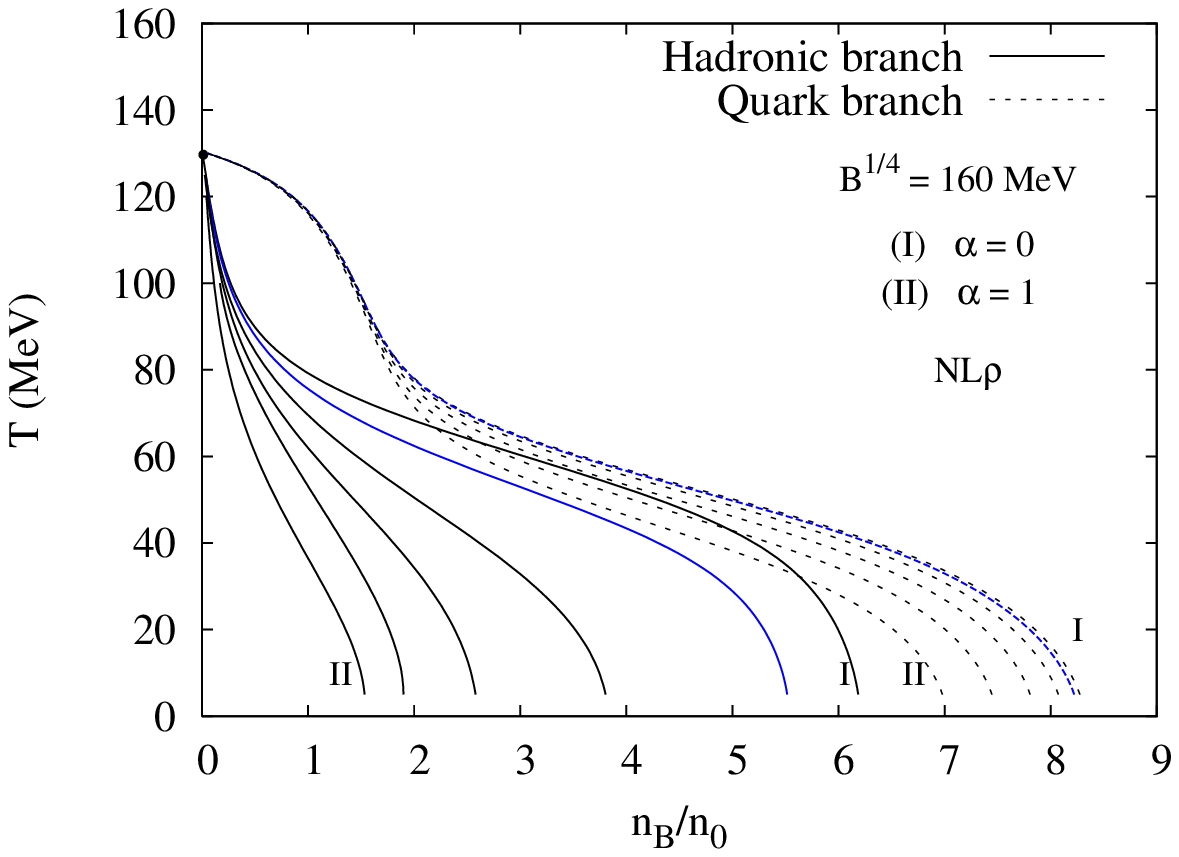}
 &
\includegraphics[width=8cm,height=6.2cm,angle=0]{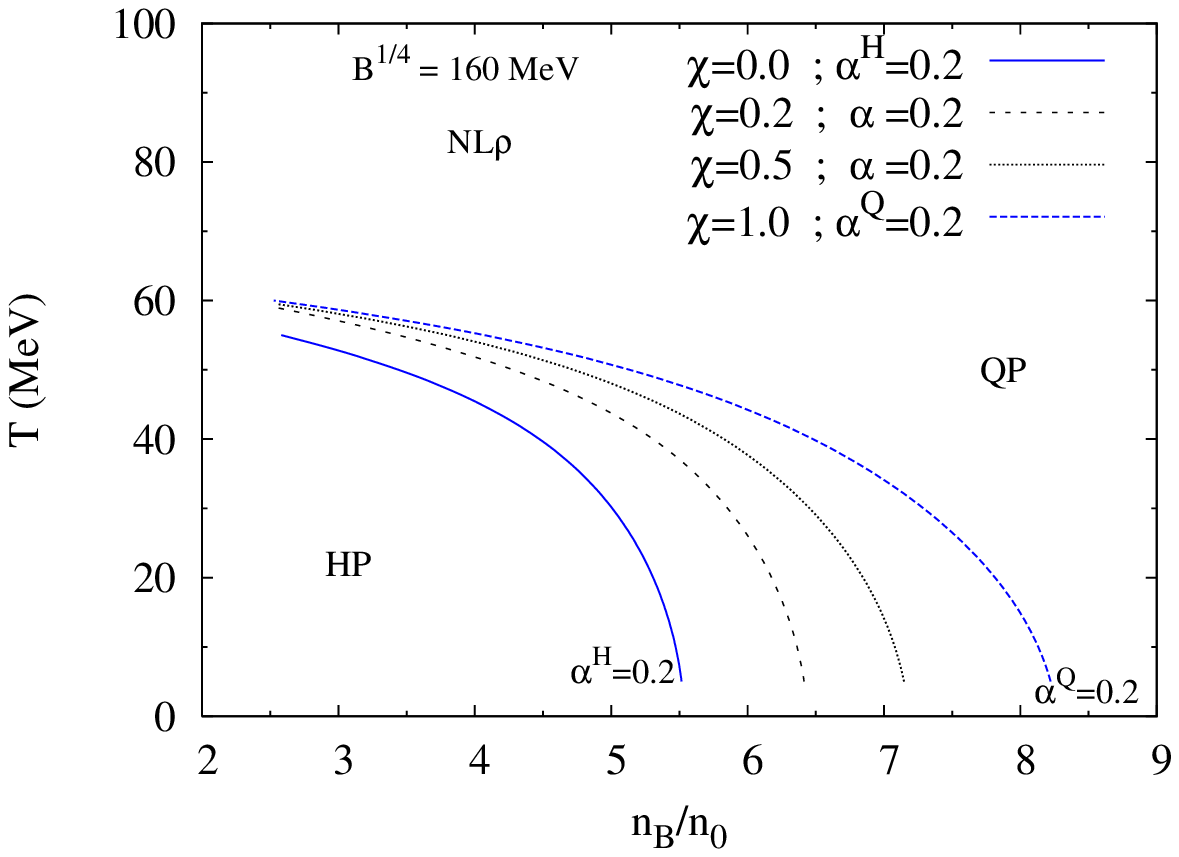}
\\ {\bf (a)} & {\bf (b)}
\end{tabular}

\caption{{\bf (a)} Same as in Fig.~\ref{temp-densb} (b) for $B^{1/4} = 160$~MeV. 
{\bf (b)} Part of Fig.~\ref{press-assym3} (a) for $\alpha = 0.2$ with the mixed phase for different 
quark concentrations ($\chi = 0.2, 0.5$).}
 \label{press-assym3}
\end{figure*}

\myindent In Fig.~\ref{temp-densb} (a) we have considered  $B^{1/4} = 190$~MeV 
together with the models NL3 and NL$\rho$. The properties of the EoS are clearly 
reflected on these results: for NL3 the transition occurs for smaller densities 
due to its very large compressibility at large densities. It is also this high 
value of the compressibility that dilutes in part the effect of 
the asymmetry parameter. NL$\rho$  has a much softer EOS and symmetry energy and, 
therefore, the curves obtained for a fixed asymmetry span a larger range of densities. 
In summary, the hadron-quark phase transition is favored when the asymmetry 
of the system is increased.

\myindent We are interested in discussing the phase transition at
intermediate temperatures and high densities and, for this reason,
we consider smaller bag pressures according
to Fig. \ref{mit-cley}.
 We set $B^{1/4} = 160$~MeV in order 
to reach a specific range of temperature and densities, which is presented in 
Figs.~\ref{press-assym-tm1} (a) and (c) and also in 
Fig.~\ref{press-assym3} (a). The asymmetries experimentally 
available up to now according to 
table~\ref{ion-asymmetries}\ref{ion-asymmetries} are in the range $0 \,-\, 0.23$.

\myindent Since NL3 is too hard and does not satisfy most of the constraints imposed by  experimental
 and observational measurements \cite{klaen06} we consider in the following 
TM1 with and
without a $\omega\rho$ non linear term which allows us to discuss a 
asy-soft and a asy-hard EoS. We also take into account 
the NL$\rho$ parametrization in order to compared with the results already 
obtained in \cite{ditoro06}.

\myindent In Fig.~\ref{press-assym-tm1} we compare TM1 and
TM1-$\omega\rho$. This allows us to discuss the effect of density dependence of the symmetry
energy on the phase transition since the isoscalar channel is kept fixed. The main effect of a
softer symmetry energy is to shift the binodal sections  for larger values of the
asymmetry parameter to larger densities. A harder symmetry energy allows the occurrence  of
the hadron-quark phase transition at smaller densities, and therefore, easier to reach with
heavy ion collisions at intermediate energies. Similar conclusion were drawn in \cite{ditoro2010} where the effect of the $\delta$-meson on the phase transition was discussed: the $\delta$-meson gives rise to a harder symmetry energy at large densities favoring the hadron-quark phase transition.

\myindent In Fig.~\ref{press-assym3} (a) we show for the same bag constant the binodal sections obtained
with NL$\rho$. It is seen that due to a softer EoS  at intermediate densities the binodal
sections occur at larger densities when compared with TM1.
The effect of the bag constant is clear if we compare Fig.~\ref{temp-densb} (b) with  $B^{1/4}
= 190$~MeV  with  Fig.~\ref{press-assym3} (a). A larger $B$ shifts the phase transition to
much larger densities, showing that in order to obtain a good estimation it is essential to
choose an adequate value of $B$.

\myindent We show in Figs.~\ref{press-assym-tm1} (b) and (d) and 
Fig.~\ref{press-assym3} (b) a part of Figs.~\ref{press-assym-tm1} (a) and (c) 
and also Fig.~\ref{press-assym3} (a), 
corresponding to $\alpha = 0.2$. We also include curves corresponding to the
mixed phase with 
the quark concentrations $\chi = 0.2$~and $0.5$ which correspond to $20\%$ and $50\%$ 
of quarks in the mixed phase. 
 One sees the indication of an interesting region 
where the phase transition probably occurs and can be probed by intermediate 
energy heavy-ion collisions. This region is located in the range 
$n_B = 2 \,-\, 4\, n_0$ and $T = 50 \,-\, 65$~MeV and can reached by the new planned 
facilities (NICA) at JINR/ Dubna \cite{nica-jinr} and (FAIR) at GSI/Darmstadt 
\cite{fair-gsi} that will start operations in the next few years.

\myindent The density behavior at intermediate/high densities defines the transition 
region. For instance,  models TM1 and TM1$\omega\rho$ would favor the detection 
of a quark phase more than NL$\rho$.

%%%%%%%%%%%%%%%%%%%%%%%%%%%%%%%%%%%%%%%%%%%%%%%%%%%%%%%%%%%%%

\section{Summary}

We have presented a study of the deconfinement phase 
transition from hadronic matter to a quark-gluon plasma, which could 
be formed in heavy-ion collisions. Calculations at finite 
temperature with a simple two-phase model and the inclusion of pion
and kaon condensation were done in order to describe this type of system. 
We have studied the effect of the
density dependence of the EoS on the phase transition choosing a convenient 
set of parametrizations of the NLWM. We have considered both hard and soft EoS 
at intermediate densities as well as models with asyhard
and asysoft symmetry energies. We have also considered the effect of gluons on the quark
phase. For the quark phase we have used the MIT bag model and chose the bag constant  
according to a parametrization of the freeze-out curve deduced from particle 
multiplicities in heavy-ion collisions \cite{mfreeze-cley}: for deconfinement 
phase transition at  $T\sim 50-60$ and  $\rho \sim 2-6\rho_0$, 
the bag constant  $B^{1/4}\sim 160$ MeV was used.

\myindent An important result is the difference between the phase diagram for a symmetric 
system and that for asymmetric matter as observed in liquid-gas phase transition. 
Usually, the onset of the phase transition takes place at lower baryon densities and 
temperatures in more asymmetric systems. This can be probed by means of neutron-rich 
nuclei in heavy-ion collisions. Moreover, the density at which the phase transition 
occurs is sensitive to the density dependence of the EoS at intermediate 
densities. A hard EoS gives rise to a transition at lower densities. 
The density dependence of the
symmetry energy also affects the transition when asymmetric matter is 
considered. The phase transition is favored for asymmetric nuclear matter, 
and even more for an asyhard symmetry energy.

\myindent  Both thermal pions and pion condensation have been included in 
the calculation. They mainly play a role at low densities, large isospin 
asymmetries and large temperatures. We have considered that the pions couple 
to the nucleons through the $\rho$-meson \cite{muller-pions}. 
Using an equivalent  parametrization for the kaon-meson coupling, which maybe too
naive  since it only takes into account the isospin interaction, we have verified that the
effect of including strangeness in the hadron phase was negligible for a system with an
overall strangeness equal to zero. This can be generalized to finite strangeness when
it becomes possible to prepare heavy-ion collisions with hypernuclei. It remains to be
investigated how sensitive are the results to the pion and kaon interaction.

\myindent The results obtained for the phase transition are very sensitive to the EOS. Both  the
isoscalar and isovector interactions have an effect on the transition density. According to 
the effective models used in this work there exists a region in the parameter space 
where the phase transition probably occurs and can be probed by heavy-ion collisions at intermediate energies. This region is located in the range 
$n_B = 2 \,-\, 4\, n_0$ and $T = 30 \,-\, 65$~MeV and can be reached by the new planned 
facilities (NICA) at JINR/ Dubna \cite{nica-jinr} and (FAIR) at GSI/Darmstadt 
\cite{fair-gsi} that will start operations in the next few years. We have obtained a larger $T$
interval, extending to lower temperatures,  for the same densities obtained in \cite{ditoro2010} due to the properties of the models
used. A clear sign of a phase transition could be used to constrain  both the EOS and symmetry energy
at intermediate densities.

\myindent We have verified that models with a soft EOS and soft symmetry energy
such as FSU do not predict a hadron-quark phase transition at densities
that could be attainned in the laboratory.

\myindent A more complete system with all baryons of the baryonic octet and strange
mesons, as well as
interacting pions and kaons,  and using  interactions constrained by experimental
measurements, is under investigation in order get more systematic results.

\section*{ACKNOWLEDGMENTS}

This work was partially supported by CNPq (Brazil) and
CAPES(Brazil)/FCT(Portugal) under project 232/09 and  by FCT (Portugal) under the grants
PTDC/FIS/64707/2006 and CERN/FP/109316/2009. 
 R. C. is grateful 
for the warm hospitality at Centro de F\'{\i}sica Computacional/FCTUC.

%\clearpage
%\newpage

\appendix

\section{}%
%\vspace{-0.5cm}
\begin{center}
\bf The boson thermodynamic potential
\end{center}
$ $
%\vspace{0.4cm}

\myindent Using the Lagrangian density in the minimal coupling scheme 
\cite{knorren-lag,schaffner-lag,prakash-rep-lag,glen-shaffner-lag,pons1,kapusta-lag}

\begin{equation}
{\cal L}_{b} = D_{\mu}^* \, \Phi^* \,\, D^{\mu} \, {\Phi} - m_{b}^{*2} \Phi^* \, {\Phi} \;,
\end{equation}
$ $

it is possible to obtain the respective thermodynamic potential and the EoS of the 
boson fields. It is convenient to 
transform $\Phi$ into real and imaginary parts using two 
real fields, $\phi_1({\bf x},t)$ and $\phi_2({\bf x},t)$ such that %$\Phi = (\phi_1 + i \, \phi_2)/2^{1/2}$

\begin{equation}
\Phi = \frac{1}{\sqrt{2}}(\phi_1 + i \, \phi_2) \;\;\; , \;\;\;  \Phi^* = \frac{1}{\sqrt{2}}(\phi_1 - i \, \phi_2) \,.
\end{equation}
$ $

The conjugate momenta are

\begin{align} 
\pi_1 &= \frac{\partial {\cal L}_b}{\partial (\partial_0 \phi_1)} = \partial_0 \phi_1 - X_0 \phi_2 \;, \nonumber \\
\\
\pi_2 &= \frac{\partial {\cal L}_b}{\partial (\partial_0 \phi_2)} = \partial_0 \phi_2 + X_0 \phi_1 \nonumber \;,
\end{align}
$ $

and the corresponding Hamiltonian density of the boson field,
${\cal H}_b = \pi_1 \partial_0 \phi_1 + \pi_2 \partial_0 \phi_2 - {\cal L}_b$~ such 
that the four-current and its zero component are
\begin{equation}
j_{\mu} = i \, \left[ \, \Phi^* (D_{\mu} \Phi) - (D_{\mu}^* \Phi^*) \Phi \, \right] \;,
\end{equation}
\begin{equation}
j_0 = \phi_2 \pi_1 - \phi_1 \pi_2 \;.
\end{equation}
$ $

For the neutral pions we just have 
$\pi = \frac{\partial {\cal L}_b}{\partial (\partial_0 \phi)} = \partial_0 \phi$~
and $\Phi = \frac{\phi}{2^{1/2}}$~, such that $\Phi^* = \Phi$ and $j_{\mu} = 0$. 
Now we can write the Hamiltonian density
\begin{widetext}
\begin{align} 
{\cal H}_b \,=\,\, &\frac{1}{2}\pi_1^2 + \frac{1}{2}\pi_2^2 + \pi_1 (X_0 \, \phi_2) 
- \pi_2 (X_0 \, \phi_1) + \frac{1}{2}({\vec \nabla} \phi_1 )^2 
+ \frac{1}{2}({\vec \nabla} \phi_2 )^2 + (\partial_i \, \phi_2) X_i \, \phi_1 \nonumber \\
\nonumber \\
&- (\partial_i \, \phi_1) X_i \, \phi_2 + \frac{1}{2}(X_i \, \phi_1)^2 
+ \frac{1}{2}(X_i \, \phi_2)^2 + \frac{m_b^{*\,2}}{2}(\phi_1^2 + \phi_2^2)  \;,
\end{align}
$ $

where $i = 1,2,3$, and the partition function in the grand canonical ensemble 
as a functional integral is given by

\begin{equation}
Z_b =  \int_{}^{}{[d \pi_1][d \pi_2] }  \int\limits_{\scriptscriptstyle \rm periodic}{[d \phi_1][d \phi_2]}\; 
\exp \left\{  \int_{0}^{\beta}{d \tau}  \int{d^3 x} \left[  i\, \pi_1 \frac{\partial \phi_1}{\partial \tau}  
+ i\, \pi_2 \frac{\partial \phi_2}{\partial \tau} 
- {\cal H}_b + \mu_b (\phi_2 \pi_1 - \phi_1 \pi_2) \right] \right\} \;,
\end{equation}
$ $

where $\mu_b$ is the boson chemical potential associated with the conserved charge 
$Q = \int d^3 x \, j_0(x)$. Here ``periodic'' means that the integration over the field 
is constrained in the imaginary time variable $\tau = i \,t$ so that 
$\phi_{k}({\bf x},0) = \phi_{k}({\bf x},\beta)$, and where $\beta = 1/T$. The 
neutral pion Hamiltonian is ${\cal H}_{\pi^0} = \frac{1}{2}\pi^2 
+ \frac{1}{2}({\vec \nabla} \phi)^2 + \frac{1}{2} m_{\pi^0}^{\,2} \phi^2$ which has 
the form of that of a neutral scalar field, so that it can be used within the relativistic 
mean field approach, as it is known that the pion pseudoscalar interaction term vanishes 
in the mean field level.
After some algebra the integration over momenta can be done and the result is

\begin{align} 
Z_b = & N^2 \int\limits_{\scriptscriptstyle \rm periodic}{[d \phi_1][d \phi_2]}\; 
\exp \left\{   \int_{0}^{\beta}{d \tau}  \int{d^3 x} \left\{ -\frac{1}{2} \left[ 
\frac{\partial \phi_1}{\partial \tau} - i\,(\mu_b - X_0)\,\phi_2  \right]^2 
-\frac{1}{2} \left[ 
\frac{\partial \phi_2}{\partial \tau} + i\,(\mu_b - X_0)\,\phi_1  \right]^2  \right. \right.   \nonumber \\
\label{func-part2} \\
& \left. {\left. -\frac{1}{2}({\vec \nabla} \phi_1)^2 - \frac{1}{2}({\vec \nabla} \phi_2)^2 
+ (\partial_i \phi_1)\,X_i \phi_2 - (\partial_i \phi_2)\,X_i \phi_1 
- \frac{1}{2} (X_i \phi_1)^2 - \frac{1}{2}(X_i \phi_2)^2 - \frac{1}{2} m_b^{*\,2}(\phi_1^2 + \phi_2^2)  % a parte de baixo eh so para fazer volume para as chaves ficarem grandes
{\mathop {}\limits_{\mathop {}\limits_{}^{}}^{\mathop {}\limits_{}^{}} } \right\}} \right\} \;, \nonumber 
\end{align}

where $N$ is a normalization factor. In the mean field approach $\langle X_i \rangle = 0$. Integrating (\ref{func-part2}) by parts, and 
taking into account the periodicity of $\phi_1$ and $\phi_2$, the result is

\begin{align} 
Z_b = & N^2 \int\limits_{\scriptscriptstyle \rm periodic}{[d \phi_1][d \phi_2]}\; 
\exp \left\{ \frac{1}{2}  \int_{0}^{\beta}{d \tau}  \int{d^3 x} \left\{ \phi_1 \left[ 
\frac{\partial^2 }{\partial \tau^2} + \nabla^2 - m_b^{*\,2} + (\mu_b - X_0)^2 \right] \phi_1 
  \right. \right.   \nonumber \\
\label{func-part3} \\
& \left. {\left.  \phi_2 \left[ \frac{\partial^2 }{\partial \tau^2} + \nabla^2 - m_b^{*\,2} + (\mu_b - X_0)^2 \right] \phi_2 
+ 2 i (\mu_b - X_0) \left[ \phi_2 \left( \frac{\partial \phi_1}{\partial \tau} \right) 
- \phi_1 \left( \frac{\partial \phi_2}{\partial \tau} \right) \right]  % a parte de baixo eh so para fazer volume para as chaves ficarem grandes
{\mathop {}\limits_{\mathop {}\limits_{}^{}}^{\mathop {}\limits_{}^{}} } \right\}} \right\} \;. \nonumber 
\end{align}

The fields can be expanded in Fourier series as

\begin{align} 
\phi_1({\bf x},\tau) = &\sqrt{2} \, \zeta \, \cos(\theta) + \left( \frac{\beta}{V} \right)^{1/2} \sum_{n}\sum_{{\bf p}} 
e^{i({\bf p}  \cdot {\bf x} + \omega_n \tau)} \phi_{1,n}({\bf p})  \;,  \nonumber \\
\label{fields-fourier1} \\
\phi_2({\bf x},\tau) = &\sqrt{2} \, \zeta \, \sin(\theta) + \left( \frac{\beta}{V} \right)^{1/2} \sum_{n}\sum_{{\bf p}} 
e^{i({\bf p}  \cdot {\bf x} + \omega_n \tau)} \phi_{2,n}({\bf p})   \;, \nonumber 
\end{align}
$ $

where the  Matsubara frequency is $\omega_n = 2 \pi n T$, due to the constraint of 
periodicity of the fields, such that $\phi_k({\bf x},\beta) = \phi_k({\bf x},0)$ 
for all ${\bf x}$. The normalization factors of (\ref{fields-fourier1}) can be chosen 
so that each Fourier amplitude is dimensionless. The infrared character of the field 
is carried out by $\zeta$ and $\theta$, so that, 
$\phi_{1,0}({\bf p}=0) = \phi_{2,0}({\bf p}=0) = 0$ which allows some particles to 
reside in the $n = 0$,~${\bf p}=0$ state, i.e., a possibility of a condensation of 
the bosons into the zero-momentum state (``s-wave'' condensation). Using (\ref{fields-fourier1}) in 
(\ref{func-part3}), and noting that $\phi_{-n}(-{\bf p}) = \phi_{n}^*({\bf p})$ because 
$\phi_1({\bf x},\tau)$ and $\phi_2({\bf x},\tau)$ are real fields, we have
\begin{equation}
 Z_b = N^2 \left[ \prod_{n} \prod_{{\bf p}} \int d\phi_{1,n}({\bf p}) \, d\phi_{2,n}({\bf p}) \right] e^S \;,
\label{func-part4}
\end{equation}
where
\begin{equation}
 S =  \beta V \zeta^2 \left[ (\mu_b - X_0 )^2 - m_b^{*\,2} \right] - \frac{1}{2} \sum_{n} \sum_{{\bf p}} \left[  \phi_{1,-n}(-{\bf p}) \;,\; \phi_{2,-n}(-{\bf p})  \mathop {}\limits_{}^{}  \right] 
 \mathbb{D}  \left[ {\begin{array}{c}
   \phi_{1,n}({\bf p})  \\
\\
   \phi_{2,n}({\bf p})  \\
\end{array}} \right] \;,
\end{equation}
$ $
and

\begin{equation}
 \mathbb{D} = \beta^2 \left[ {\begin{array}{cc}
   \omega_n^2 + {\bf p}^2 + m_b^{*\,2} - (\mu_b - X_0)^2 \;\; &  - 2 (\mu_b - X_0) \omega_n  \\
 & \\
   2 (\mu_b - X_0) \omega_n  & \;\; \omega_n^2 + {\bf p}^2 + m_b^{*\,2} - (\mu_b - X_0)^2 \\
\end{array}} \right] \;.
\end{equation}

As the thermodynamic potential is given by $\Omega = - (1/\beta) \ln (Z)$, we can perform 
the integrals in (\ref{func-part4}) and write

\begin{equation}
\ln (Z_b) = \beta V \zeta^2 \left[ (\mu_b - X_0 )^2 - m_b^{*\,2} \right] 
+ \ln \left[ ({\rm det} \, \mathbb{D} )^{-\frac{1}{2}} \right] \;.
\label{ln-det-zb}
\end{equation}
$ $

The multiplication of $Z_b$ by any constant is irrelevant since it does not 
change the thermodynamics of the system. The second term of (\ref{ln-det-zb}) is given by

\begin{align} 
- \frac{1}{2} \ln \left[ {\rm det}\, \mathbb{D} \right] &= - \frac{1}{2} \ln \left\{ \prod_{n} 
\prod_{\bf p} \beta^4 \left[  \left( \; \omega_n^2 + {\bf p}^2 + m_b^{*\,2} - (\mu_b - X_0)^2 \; \right)^2 + 4 (\mu_b - X_0)^2 \omega_n^2 \; \right] \right\}  \nonumber \\
 \\
&= - \frac{1}{2} \ln \left\{ \prod_{n,{\bf p} } \beta^2 \left[  \omega_n^2 + (\omega^+ - \mu_b)^2  \right]  \right\}  
- \frac{1}{2} \ln \left\{ \prod_{n,{\bf p} } \beta^2 \left[  \omega_n^2 + (\omega^- + \mu_b)^2  \right]  \right\}   \;, \nonumber 
\end{align}

so that (\ref{ln-det-zb}) can be written as

\begin{equation}
\ln (Z_b) = \beta V \zeta^2 \left[ (\mu_b - X_0 )^2 - m_b^{*\,2} \right] 
-\frac{1}{2} \sum_{n,{\bf p}} \ln \left\{ \beta^2 \left[  \omega_n^2 + (\omega^+ - \mu_b)^2  \right]  \right\} 
-\frac{1}{2} \sum_{n,{\bf p}} \ln \left\{ \beta^2 \left[  \omega_n^2 + (\omega^- + \mu_b)^2  \right]  \right\}    \;,
\end{equation}

and in the continuum limit, neglecting the zero-point energy contribution, due to the 
mean field approach, the result is
\begin{equation}
\ln (Z_b) = \beta V \zeta^2 \left[ (\mu_b - X_0 )^2 - m_b^{*\,2} \right] 
- V \int \frac{d^3 p}{(2 \pi)^3}  \left\{ \; \ln \left[ 1 - e^{-\beta(\omega^+ - \mu)}  \right] 
+  \ln \left[ 1 - e^{-\beta(\omega^- + \mu)}   \right] \; \right\} \;,
\end{equation}
where 
\begin{equation}
\omega^{\pm}(p) \equiv \sqrt{p^2 + m_b^{*\,2}} \; \pm \; X_0 \;,
\end{equation}
$ $

is the effective Bose energy, such that the thermodynamic potential for the 
bosons is given by

\begin{equation}
\frac{\Omega_b}{V} = - \frac{\ln (Z_b)}{\beta V} = \zeta^2 \left[ m_b^{*\,2} - (\mu_b - X_0 )^2 \right] 
+ T \int \frac{d^3 p}{(2 \pi)^3}  \left\{ \; \ln \left[ 1 - e^{-\beta(\omega^+ - \mu)}  \right] 
+  \ln \left[ 1 - e^{-\beta(\omega^- + \mu)}   \right] \; \right\} \;.
\label{therm-pot-boson}
\end{equation}
$ $

\end{widetext}

\end{document}